\documentclass[10pt,aps,amsfonts,amssymb,floatfix,longbibliography,superscriptaddress,notitlepage,twocolumn,footinbib]{revtex4-1}
\usepackage{mathrsfs} 
\usepackage{graphicx,color}
\usepackage{bm}
\usepackage[normalem]{ulem}
\usepackage{physics}
\usepackage{afterpage}
\usepackage{ulem}
\usepackage[colorlinks=true]{hyperref}
\hypersetup{
  colorlinks=true,
  linkcolor=blue,
  filecolor=magenta,
  citecolor=blue,
  urlcolor=blue,
}

\def\ol{\overline}

\def\to{\rightarrow} \def\la{\left\langle} \def\ra{\right\rangle}

\newcommand{\red}[1]{\textcolor{black}{#1}}
\newcommand{\eq}[1]{Eq.~(\ref{#1})}	
\newcommand{\Sec}[1]{Sec.~\ref{#1}}
\newcommand{\Figcaption}[1]{\def\@captype{figure}\caption{#1}}
\newcommand{\tblcaption}[1]{\def\@captype{table}\caption{#1}}
\makeatother

\newcounter{num}

\newcommand{\ctext}[1]{\raise0.1ex\hbox{\scriptsize \textcircled{\tiny {#1}}}}
\newcommand{\cctext}[1]{\raise0.1ex\hbox{\textcircled{\scriptsize {#1}}}}

\newcommand{\beq}{\begin{equation}}
\newcommand{\eeq}{\end{equation}}

\newcommand{\gone}{\rm \hspace{.18em}i\hspace{.18em}}
\newcommand{\gtwo}{\rm \hspace{.08em}ii\hspace{.08em}}
\newcommand{\gthree}{\rm i\hspace{-.08em}i\hspace{-.08em}i}
\newcommand{\gfour}{\rm i\hspace{-.08em}v\hspace{.06em}}
\newcommand{\gfive}{\rm \hspace{.06em}v\hspace{.06em}}
\newcommand{\gsix}{\rm \hspace{-.06em}v\hspace{-.08em}i}

\begin{document}
\title{Supercooled Jahn-Teller Ice}

\author{Kota Mitsumoto}
\affiliation{Molecular Photoscience Research Center, Kobe University, Kobe 657-8501, Japan}

\author{Chisa Hotta}
\affiliation{Department of Basic Science, University of Tokyo, Tokyo 153-8902, Japan}

\author{Hajime Yoshino}
\affiliation{Cybermedia Center, Osaka University, Toyonaka, Osaka 560-0043, Japan}
\affiliation{Graduate School of Science, Osaka University, Toyonaka, Osaka 560-0043, Japan}

\begin{abstract}
When the spins on the frustrated pyrochlore lattice obey the celebrated 2-{\it in}-2-{\it out} ice rule, they stay in a correlated disordered phase and break the third law of thermodynamics. Similarly, if the atomic ions on the pyrochlore lattice move in and outward of the tetrahedra, they may obey a constraint resembling the ice rule. We discover that a model for pyrochlore molybdates $A_2$Mo$_2$O$_7$ ($A=$Y, Dy, Tb) exhibits a ``supercooled ice" state of the displacement degrees of freedom of Mo$^{4+}$ ions, when we take account of the Jahn-Teller (JT) effect. The JT effect occurs when the lattice distortions reduce the symmetry of the local crystal field, resulting in the orbital-energy-splitting that causes the local energy gain. Unlike the standard JT effect that leads to periodic long range ordering, the displacements of Mo$^{4+}$ ions are disordered following the ice-like rule. We microscopically derive a model that describes this situation by having the 2nd and 3rd neighbor interactions between {\it in}-{\it out} lattice displacements comparably as strong as the nearest neighbor interactions of standard ice. There, the well-known nearly flat energy landscape of the ice state is altered to a metastable highly quasi-degenerate ice-like liquid state coexisting with a crystalline-like ground state. Our Monte Carlo simulations show that this liquid remains remarkably stable down to low temperatures by avoiding the putative first order transition. The relaxation in the supercooled JT ice state exhibits glassy dynamics with a plateau structure. They fit the feature of a ``good glassformer" very often found in molecular liquids, but that has never been observed in material solids. The high glass-forming ability of the interacting lattice degrees of freedom will play a key role in the spin-glass transition of the material.
\end{abstract}

\maketitle

\begin{figure*}[t]
\includegraphics[width=160mm]{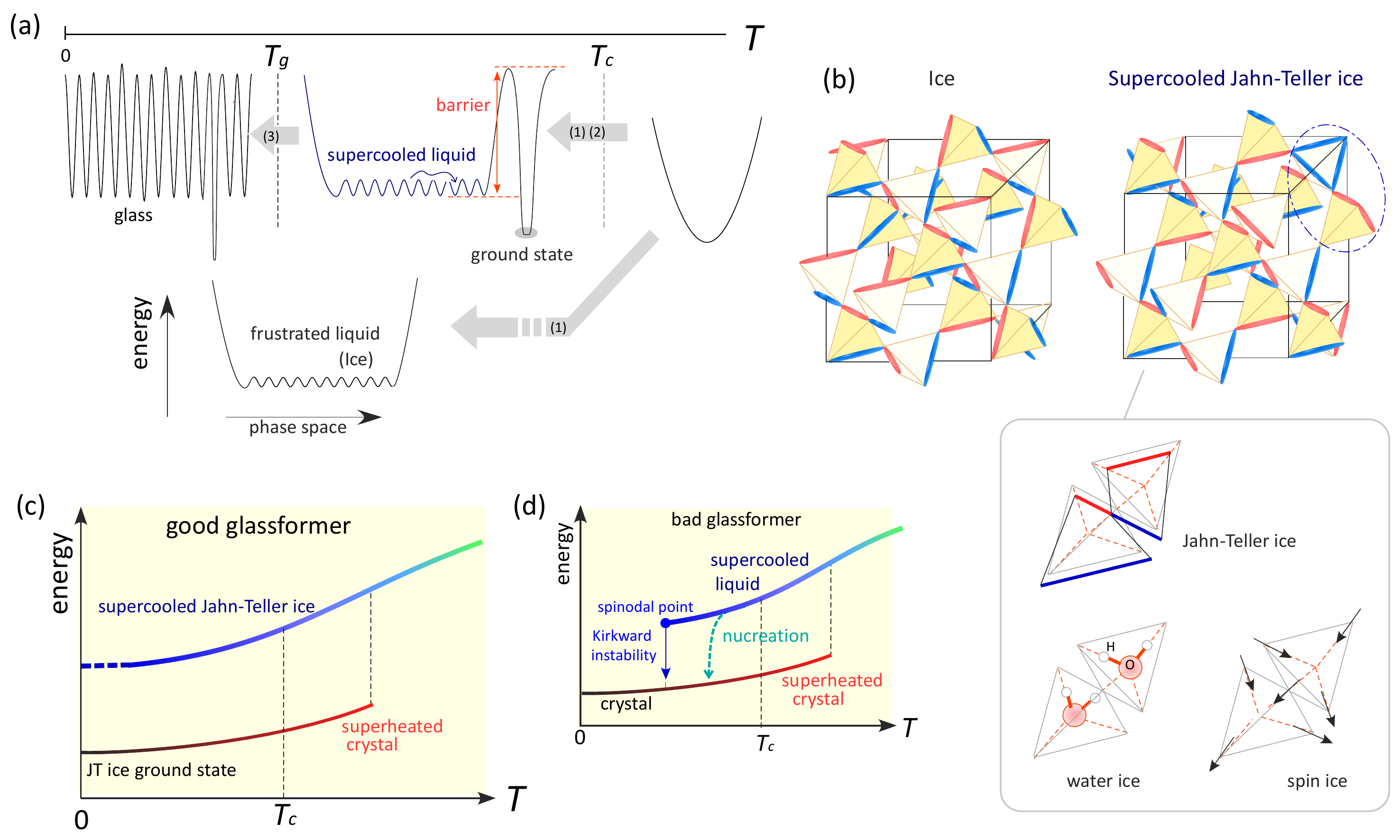}
\caption{
(a) Schematic energy landscapes for various temperatures; the single free energy minimum will develop into a coexisting metastable supercooled liquid state and a crystalline solid ground state below the first order transition temperature, $T_{\rm c}$. When the energy barrier is high we call the supercooled liquid a `good glass former''. In lowering the temperature the relaxation time (viscosity for a supercooled molecular liquid) increases rapidly in accordance with the development of energy barriers between multivalleys called ``basins"  toward the glass phase. This glassforming picture is typical of molecular liquids. For frustrated systems on a lattice (materials), the system gradually crosses over to the quasi degenerate low energy landscape which remains disordered toward zero temperature. 
(b) Ice and supercooled JT ice state on a pyrochlore lattice. Lower panels show an ice rule for a water ice, spin ice, and supercooled JT ice. Red and blue bonds indicate the $in$-$in$ and $out$-$out$ configurations of spins or lattices. The supercooled JT ice consists of bonds that are bent at each vertex, and partially breaks the 2-$in$-2-$out$ rule; e.g., 3-$in$-1-$out$ or 1-$in$-3-$out$ tetrahedra are inserted as monopoles marked in an oval that is not easy to move and is not easily annihilated. 
(c,d) Schematic energy diagrams of `good" and `bad glass-forming" molecular liquid systems as functions of temperature. There is a coexistent metastable liquid state and the crystalline ground state for finite temperature range, and the true thermodynamic first-order transition between the two takes place at a particular $T_c$. During the cooling process, the thermally activated nucleation may lead to the transition from the supercooled liquid to the crystalline state. 
}
\label{fig_supercooling}
\end{figure*}
\section{Introduction}
The glass-forming liquid is an intriguing state of matter which has a mixed character of solids and liquids \cite{angell2000relaxation}. At a short timescale, it behaves as a glass, but at a much longer timescale, it flows as a liquid slowly but unboundedly. There are numerous examples of glass-forming liquids made of molecules, polymers and colloids; they are the metastable supercooled liquid states, and coexist in the energy landscape with the thermodynamically stable crystalline solid state (see Fig.~\ref{fig_supercooling}(a)). In lowering the temperature, the timescale to relax among the metastable basins increases rapidly and the system eventually behaves as a glass. However, during the cooling process, the supercooled liquids can easily transit into the crystalline ground state. Therefore, a stable supercooled liquid that can kinetically avoid the first-order transition, even by relatively slow cooling, is called a ``good glass-former". 
\par
This prototype glass-forming picture\cite{KTW89,debenedetti2001supercooled,BB09} is established by the mean-field theory in the large dimensional limit\cite{parisi2020theory}, where they predict a complex free-energy landscape of supercooled liquids with exponentially large numbers of coexisting multi-valleys, which is believed to undergo a Kauzmann transition\cite{Ka48} selecting one of them. 
Whereas, the validity of the prediction in a realistic three-dimensional space is currently under debate, and understanding the microscopic principles of generating such energy landscape 
is an outstanding unsolved problem of physics.
\par
Solid materials can be an ideal platform to tackle this issue as they provide possibilities to investigate various liquid states based on their constituent charges, orbitals, spins and lattice distortions. Unfortunately, however, glass-forming liquid states have been scarcely observed among them. In this work, we find a microscopic model for Jahn-Teller distortions of pyrochlore molybdates $A_2$Mo$_2$O$_7$ ($A=$Y, Dy, Tb) that behaves as an ideally ``good glass-former". Although the material is known to host a spin glass phase, they are clean and do not fit to the spin glass picture that requires a large amount of randomness\cite{mydosh1993spin,saunders2007spin,shinaoka2011spin}. Our supercooled liquid picture in a solid material serves as a basis of a glass-forming mechanism that can be directly compared with molecular liquids. 
\par
A key factor is a {\it frustration} \cite{tarjus2011overview} that provides three indispensable conditions to enable glass-forming liquids; (1) it allows the liquid state to remain (meta)stable down to low temperatures, (2) it enhances the glass forming ability or the ability to avoid crystallization kinetically\cite{cavagna2009supercooled}, and (3) it enables creation of complex free-energy landscape with competing multi-valleys\cite{KTW89,debenedetti2001supercooled,BB09,parisi2020theory}. Since frustration can be found in a broader range of systems other than the molecular, polymer, and colloidal liquids, a natural question arises: can we create a glass forming liquid made of other constituents? Spins on the geometrically frustrated lattice can afford condition (1); in the spin-ice state\cite{bramwell2020history, ramirez1999zero} of the pyrochlore lattice antiferromagnets, the strong local constraints called ``ice rule" arising from the frustration will select the disordered configurations down to zero temperature\cite{gardner2010magnetic}. However, {\it it is not a glass-forming liquid}, since their energy landscape is essentially flat except for small energy barriers(see Fig.~\ref{fig_supercooling}(a)), satisfying neither condition (2) nor (3). 
\par
A pyrochlore lattice consists of corner-sharing tetrahedra as shown in Fig.~\ref{fig_supercooling}(b). In each tetrahedron, the ice rule forces the spins to have a 2-{\it in}-2-{\it out} configuration, similarly to the water-ice whose two hydrogen atoms placed at the vertices of the tetrahedra move toward and the other two away from the center oxygen as shown in Fig. \ref{fig_supercooling}(a). The spin ice has a large degeneracy that breaks the third law of thermodynamics\cite{pauling1935structure}, while, unlike glasses, they can transform from one to another by exciting a monopole and moving them without an energy barrier, which itself is a source of exotic U(1) spin liquids\cite{hermele2004u1sl,lee2012u1sl}. 
\par
Recently, another possibility beyond the spin ice picture has been suggested experimentally in pyrochlore molybdate Y$_2$Mo$_2$O$_7$ that the lattice displacements of the Mo$^{4+}$ ($4d^2, S=1$) ion may be disordered and follow a 2-{\it in}-2-{\it out} ice rule \cite{thygesen2017orbital}. Later on, the underlying mechanism of disorder is speculated as some sort of a Jahn-Teller (JT) effect \cite{thygesen2017orbital, smerald2019giant, mitsumoto2020spin}. 
\par
The conventional JT effect takes place as a local energy optimization process, which can be simply repeated in space to optimize the global energy\cite{JT1937,oBrien1993,goodenough1998}: when the orbitals of a single ion have a degeneracy due to some local symmetries of the crystal field from its surroundings, the electrons occupying them can gain energy by the lattice displacements that lower the crystal-field-symmetry and lift the orbital degeneracy. 
This competes with the increase of the elastic energy, and an energetically optimal, finite lattice displacement occurs. 
The JT distortion is usually the same for all ions since the local JT energy gains of different ions are determined independently of each other. However, in the pyrochlore lattice, 
it turns out that such locally optimal JT energies associated with different ions conflict with each other, so those lattice displacements cannot be determined solely by a local crystal field on a single ion. 
\par
We describe this correlated JT effect unbiasedly in the microscopic Ising model, which has the second and third nearest-neighbor interactions, comparably as strong as the nearest neighbor ones in the ice model. These second and third neighbor interactions impose a tougher restriction to the 2-$in$-2-$out$ state. Resultantly, the flat energy landscape of an ice is altered, developing the multivalley and high barrier referred to in conditions (2) and (3) for a glassformer (see the one in Fig.~\ref{fig_supercooling}(a)). 
This landscape is anticipated from the Monte Carlo simulations, indicating two competing phases; 
one is the lowest energy crystalline-like state consisting of 2-$in$-2-$out$ displacements obeying the bending ice rule. 
The other is a supercooled disordered state where lattice distortions of 90\% of the tetrahedra have the 2-$in$-2-$out$ state that obeys the bending ice rule. 
Most importantly, the supercooled liquid turns out to be remarkably stable to be regarded as a good glassformer, possibly having a very high energy barrier toward crystallization; 
we call it a ``supercooled JT ice" (see Fig.~\ref{fig_supercooling}(c)). 
\par
As a matter of fact, supercooled liquids are rarely found in systems other than {\it molecular} liquids. The charge glass phase in the organic $\theta$-ET$_2X$ crystal exceptionally exhibits a supercooled liquid behavior, but it transits into a long-range ordered phase on slowly cooling the sample\cite{kagawa2013charge}. Two mechanisms are known to destabilize supercooled liquids as shown in Fig.~\ref{fig_supercooling}(d); one is the Kirkwood instability which is the local instability of a liquid state toward crystallization \cite{kirkwood1951phase, klein1986instability, cugliandolo2020mean}. The other is the nucleation process of the crystal phase from the liquid phase \cite{cavagna2009supercooled}. In principle, the latter cannot be avoided since the free energy of the crystalline state is lower than that of the liquid state. In theory, a supercooled paramagnetic phase coexisting with long-ranged orders such as ferromagnets is conceivable, but only proved to exist in large-dimensional (mean-field) frustrated models \cite{franz2001ferromagnet,yoshino2018disorder,cugliandolo2020mean}. 
\par
Supercooled liquids can be a source of glass. Indeed, we have previously shown that a thermodynamic glass transition in a model similar to our JT ice Hamiltonian becomes a glass\cite{mitsumoto2020spin}; there we take account of the $S=1$ spin degrees of freedom of Mo$^{4+}$ ions and couple them with the JT ice degrees of freedom. This explains the disorder-free spin-glass transition of $A_2$Mo$_2$O$_7$ ($A=$Y, Dy, Tb) \cite{greedan1986spin, gaulin1992spin, dunsiger1996muon, gingras1997static, gardner1999glassy, hanasaki2007nature}. The supercooled JT ice without quenched disorder is thus a naturally arising phenomenon in a clean three-dimensional bulk crystalline solid, which had not been conceived both in theories and in experiments. 
\par
The paper is organized as follows. 
In \Sec{sec_elastic} and \Sec{sec_jti} we analyze the elastic and JT energies of electrons on Mo$^{4+}$ ions on the pyrochlore lattice when the lattice displacement of Mo$^{4+}$ ions takes place, finding that the softest low energy modes move the Mo$^{4+}$ ions {\it in} and {\it out} of the unit tetrahedron. 
The {\it in-out} displacements change the relative angle of Mo-O bonds and lower the crystal field symmetry, and the resultant JT energy gains are associated with a variety of types of {\it in-out} displacements. 
In \Sec{sec_simplify} we derive an effective JT ice model based on the energetics obtained in the former sections. 
We derive an effective unbiased Ising-type Hamiltonian that reproduces the microscopic energetics of the JT distortions, 
which reveals that the {\it in-out} displacements of several Mo$^{4+}$ ions on the pyrochlore lattice 
interact with each other. 
In ~\Sec{sec-MC} we perform a Monte Carlo simulation on the JT ice model, 
and discover a supercooled liquid behavior. 
The relevance to the material systems and the implications of the present results are discussed in the final section. 

\begin{figure}[t]
\includegraphics[clip,width=70mm]{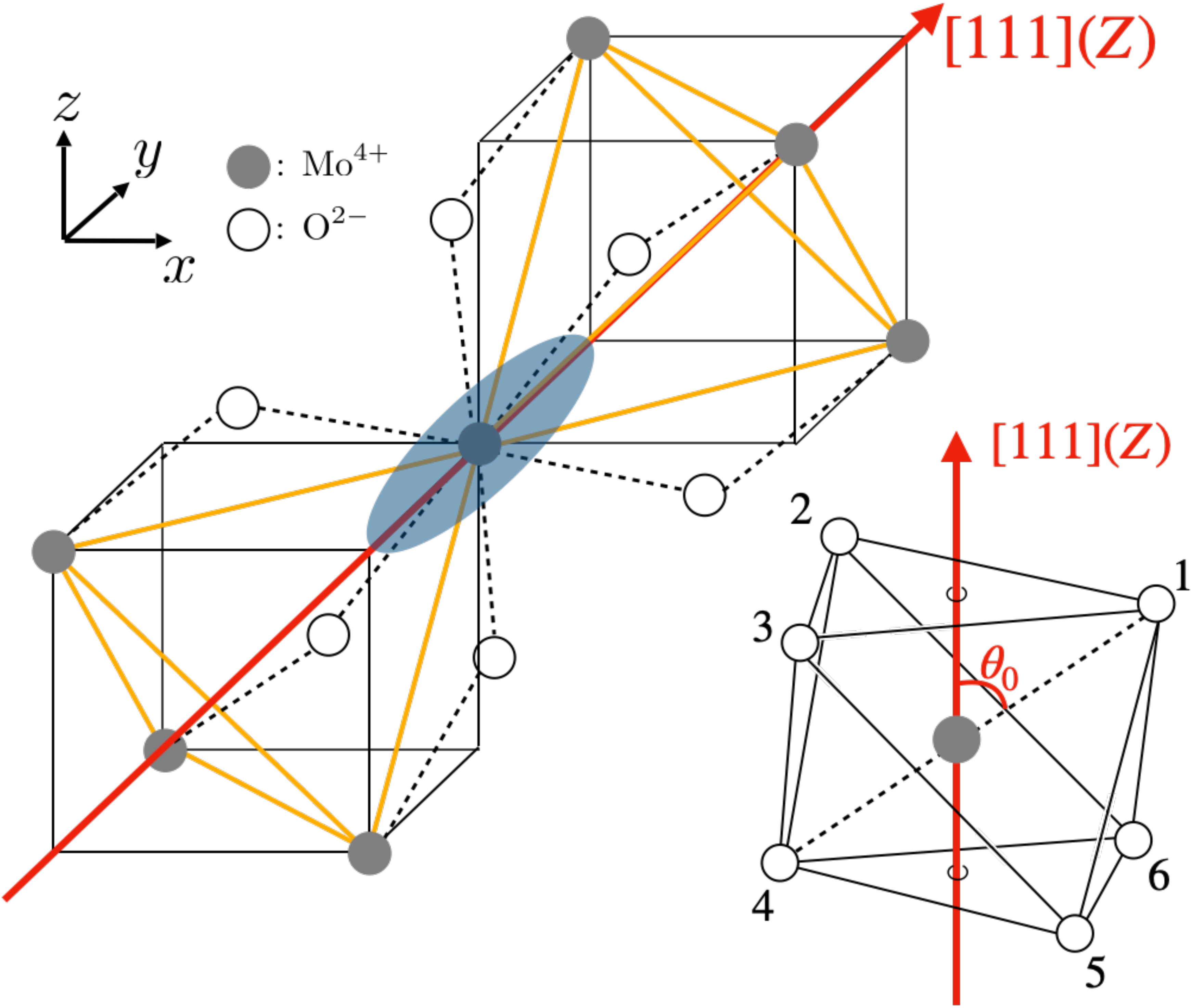}
\caption{
Local structure around a Mo$^{4+}$ ion in $A_2$Mo$_2$O$_7$.
Shaded and open circles represent Mo$^{4+}$ and O$^{2-}$ ions, respectively. 
The ellipsoid on the center Mo$^{4+}$ ion represents the elastic potential from the surrounding O$^{2-}$ ions
(See \eq{eq-v1}).
We take the {\it local} $Z$-axis in the [111] direction with its origin at the Mo$^{4+}$ ion. 
The angle $\theta_0$ of Mo-O bond about the $Z$-axis determines 
both the mechanical property and the orbital energy level splitting due to the trigonal crystal field. 
}
\label{trigonal}
\end{figure}
\begin{figure}[t]
\includegraphics[clip,width=75mm]{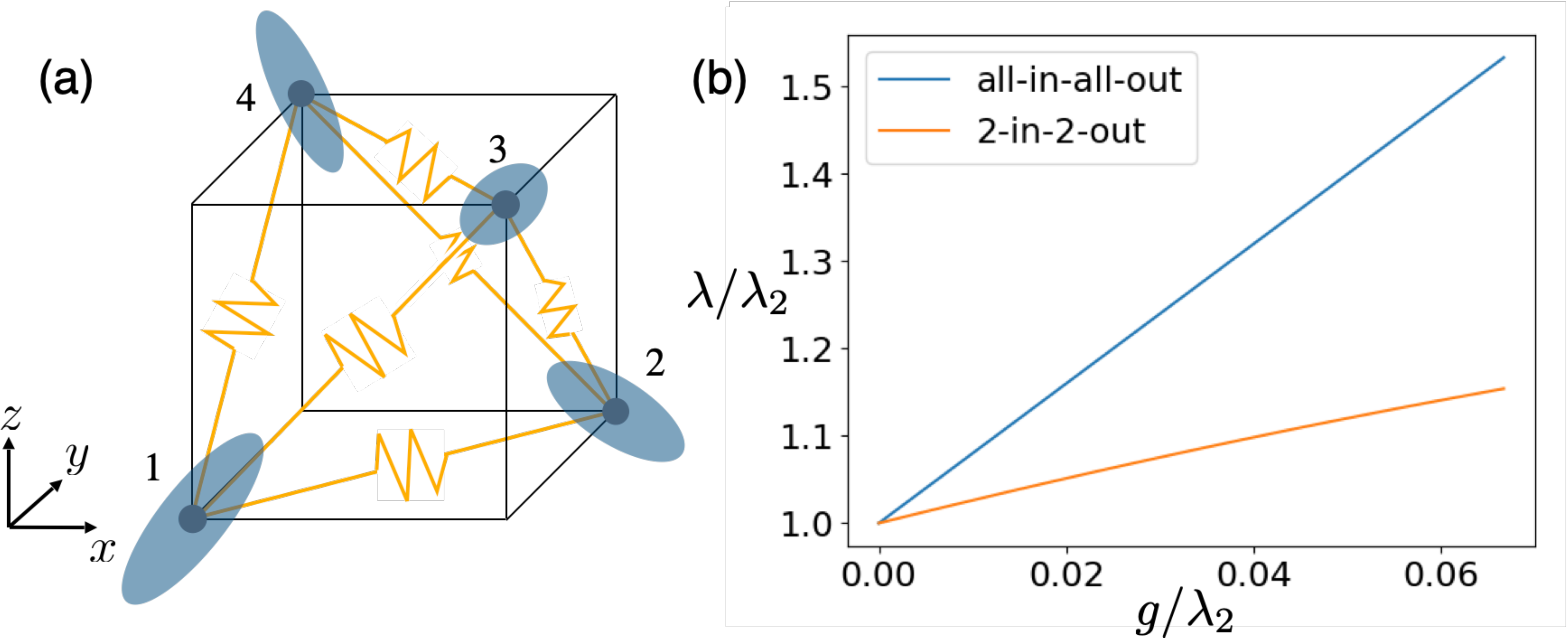}
\caption{
(a) One-body ($V_1$) and two-body ($V_2$) potentials on single tetrahedron represented by the ellipsoids and zigzag bonds, respectively.
(b) Lowest eigenvalue $\lambda$ of the Hessian matrix in Eq.(\ref{eq:Hessian}) including the two-body vibrational energy scaled by the coupling constant $g$ of $V_2$. 
At $g\ne 0$ the four-fold degeneracy of the lowest energy, $\lambda_2$, of the one-body potential 
are lifted to the three 2-$in$-2-$out$ modes and a single all-$in$-all-$out$ mode. 
}
\label{tetra}
\end{figure}
\section{Elastic energy} 
\label{sec_elastic}
In this section, we evaluate the elastic energy loss due to lattice displacements. As shown in Fig.~\ref{trigonal}, each Mo$^{4+}$ is octahedrally coordinated by six oxygen ions O$^{2-}$ which forms a trigonal crystal field, since among the eight faces of octahedra, a pair of triangles facing each other in the [111] direction of the figure is closer than the other pairs  \cite{reimers1988crystal, gardner1999glassy, solovyev2003effects}. 
It is thus convenient to choose the $Z$-axis along the [111] direction. 
The positions of the ligands $\bm{R}_i = (X_i,Y_i,Z_i)~(i=1,2,...,6)$ can be written as 
\beq
\bm{R}_i = a_{1}(\sin \theta_i \cos \phi_i, \sin \theta_i \sin \phi_i, \cos \theta_i)
\label{trigo1}
\eeq
where $a_1 = a_{\text{Mo-O}}$ is the lattice constant of the Mo-O bond,
\beq
\begin{split}
\theta_1 = \theta_2 = \theta_3 &= \theta_0, \\
\theta_4 = \theta_5 = \theta_6 &= \pi - \theta_0, \\
\end{split}
\label{trigo2}
\eeq
 and 
\beq
\begin{split}
\phi_1 &= 0, ~\phi_2 = 2\pi/3, ~\phi_3 = 4\pi/3,  \\
\phi_4 &= \pi, ~\phi_5 = 5\pi/3, ~\phi_6 = \pi/3.
\end{split}
\label{trigo3}
\eeq
Here, $\theta_0$ is the only parameter that
controls the geometric and mechanical properties of the system.

\subsection{Single ion}
Now, to obtain the vibrational eigenmode of a single Mo$^{4+}$ we consider a two-body potential $v(|\bm{R}-\bm{R_i}|)$ which only depends on the distance between the Mo$^{4+}$ and the ligand.
For the system to be stable against (de)compression, 
the second derivative $v''(a_1)$ must be positive.
\par
For the fixed $\bm{R}_i$, one can regard
\beq
V_1(\bm{R}) = \sum_{i=1}^6 v(|\bm{R}-\bm{R_i}|)
\label{eq-v1}
\eeq
as a one-body potential, and then the $3 \times 3$ Hessian matrix of the potential is given by,
\beq
H^{\mu \nu} = \frac{\partial^2 V_1(\bm{R})}{\partial X^\mu \partial X^\nu},
\eeq
where $X^\mu, X^\nu  = X,Y,Z$.
We obtain two-fold degenerate eigenvalues $\lambda_1$ and non-degenerate eigenvalue $\lambda_2$ as
\begin{align}
\lambda_1 &= (-3\sin^2\theta_0 + 6)\frac{v'(a_1)}{a_1} + 3\sin^2 \theta_0 v''(a_1), \label{lambda_1} \\
\lambda_2 &= (-6\cos^2\theta_0 + 6)\frac{v'(a_1)}{a_1} + 6\cos^2 \theta_0 v''(a_1).\label{lambda_2}
\end{align}
The doubly-degenerate eigenmodes corresponding to $\lambda_1$ are 
confined to the displacements within the $X$-$Y$ plane, 
and non-degenerate eigenmodes corresponding to $\lambda_2$ represent the stretching along the $Z$-axis.
Using $\lambda_1$ and $\lambda_2$, the one-body potential can be written in an ellipsoidal form,
\beq
V_1(\bm{R}) = \frac{1}{2}\lambda_1\qty(X^2 + Y^2) + \frac{1}{2}\lambda_2 Z^2.
\label{ellips}
\eeq
In Fig.~\ref{trigonal}, the shaded ellipsoid describes the potential. 
\par
In the case of the regular octahedron with $\theta_0^{\rm oct} = \cos^{-1}(1/\sqrt{3}) \approx 54.74^\circ$, $\lambda_1$ is equal to $\lambda_2$.The value of $\theta_0$ of the pyrochlore spin glass $A_2$Mo$_2$O$_7$ is much larger than $\theta_0^{\rm oct} $, e.g. $\theta_0 \approx 61.76$ in the case of Y$_2$Mo$_2$O$_7$ \cite{reimers1988crystal, gardner1999glassy, solovyev2003effects}.
Therefore, from \eq{lambda_1} and \eq{lambda_2}, we can easily find that $\lambda_2$ is smaller than $\lambda_1$ if 
\beq
\frac{v'(a_1)}{a_1} < v''(a_1).
\label{condition_tri}
\eeq
Let us suppose that $v(r)$ can be represented by the 12-6 Lennard-Jones potential as
\beq
v(r) = 4\epsilon \qty[\qty(\frac{\sigma}{r})^{12} - \qty(\frac{\sigma}{r})^{6}],
\eeq
where $\epsilon$ is the energy scale of the potential and $\sigma$ 
can be regarded as the size of the ion.
In this case, $\sigma$ corresponds to the sum of ionic radii of Mo$^{4+}$ and O$^{2-}$.
The equilibrium position $r_0$ which satisfies 
$v'(r_0) = 0$
is $r_0 = 2^{1/6}\sigma$.
If $r>r_0$, the force between ions is attractive, and otherwise repulsive. 
To satisfy \eq{condition_tri}, $a$ should be smaller than $(7/2)^{1/6}\sigma$.
In the case of Y$_2$M$_2$O$_7$, the ionic radii of Mo$^{4+}$ and O$^{2-}$ are 0.65 (\AA) and 1.38 (\AA), respectively, and their sum 2.03 (\AA) is comparable with the lattice constant $a_1=2.03$ (\AA), which fulfills the above condition \cite{shannon1976revised, thygesen2017orbital}. 
We find $\lambda_1/\lambda_2 \simeq 1.94$, which means that the lattice displacement along the $Z$-axis is the softest. 
Intuitively, this is because the direction corresponding to $\lambda_2$ has larger spacing to avoid the repulsive force $v(a)$ from
the
ligands.
\subsection{Single tetrahedron}

Next we consider the vibrational mode of a single tetrahedron of Mo$^{4+}$ ions as shown in Fig. \ref{tetra} (a).
The positions of Mo$^{4+}$ ions are $\bm{r}_i = (x_i,y_i,z_i)~(i=1,2,3,4)$ and their equilibrium positions $\bm{r}_i^0 = (x_i^0,y_i^0,z_i^0)$ are 
\beq
\begin{split}
\bm{r}_1^0& = a_2/\sqrt{2}(0,0,0), ~~~\bm{r}_2^0 = a_2/\sqrt{2}(1,1,0), \\
\bm{r}_3^0& = a_2/\sqrt{2}(1,0,1), ~~~\bm{r}_4^0 = a_2/\sqrt{2}(0,1,1),
\end{split}
\eeq
where $a_2 = a_{\text{Mo-Mo}}$ is the lattice constant of the Mo-Mo bond and it is given by 
\beq
a_2 = a_1\qty(\sqrt{\frac{2}{3}}\cos\theta_0 - \sqrt{\frac{1}{3}}\sin\theta_0).
\eeq
We consider a correction to the elastic energy due to the two-body interaction $V_2(|\bm{r}-\bm{r}'|)$ between Mo$^{4+}$ ions 
in addition to the one-body potential $V_1(\bm{R})$ given by \eq{eq-v1}.
Here, the second derivative $V_2''(a_2)$ should be positive to keep the equilibrium position stable. 
For simplicity, we suppose $V_2'(a_2) = 0$ although the case of non-zero $V_2'(a_2)$ leads to qualitatively the same result.
The total energy of a tetrahedron is given by,
\beq
E_{\rm vib}(\{\bm{r}_i  \}) = \sum_{i=1}^4 V_1(\bm{R}_i) + g \sum_{i<j} V_2(|\bm{r}_i-\bm{r}_j|),
\label{vib_ene}
\eeq
where $g$ is a small parameter that represents the energy scale of $V_2$ 
and $\bm{R}_i = (X_i,Y_i,Z_i)$ is
\beq
\mqty(X_i \\ Y_i \\ Z_i) = \mqty(-\frac{1}{\sqrt{6}} 
   & -\frac{1}{\sqrt{6}} & \frac{2}{\sqrt{6}} \\
	\frac{1}{\sqrt{2}} & -\frac{1}{\sqrt{2}} & 0 \\
	\frac{1}{\sqrt{3}} & \frac{1}{\sqrt{3}} & \frac{1}{\sqrt{3}})\mqty(\hat{x}_i  \\ \hat{y}_i \\ \hat{z}_i )
\label{Z_axis}
\eeq
with 
\beq 
\begin{split}
\hat{x}_i &= (a_2/\sqrt{2}-2x_i^0)(x_i - x_i^0), \\
\hat{y}_i &= (a_2/\sqrt{2}-2y_i^0)(y_i - y_i^0), \\
\hat{z}_i &= (a_2/\sqrt{2}-2z_i^0)(z_i - z_i^0).
\end{split}
\eeq
The $12\times12$ Hessian matrix is given by,
\begin{align}
\nonumber
H^{\mu\nu}_{ij} &=  \frac{\partial^2 E_{\rm vib}(\{\bm{r}_i\})}{\partial x_i^\mu \partial x_j^\nu} \\ \nonumber
&= \delta_{ij} \qty(\frac{\partial^2 V_1(\bm{R}_i)}{\partial x_i^\mu \partial x_i^\nu} 
+ g\sum_{j \neq i} \frac{\partial^2 V_2(|\bm{r}_i-\bm{r}_j|)}{\partial x_i^\mu \partial x_i^\nu}) \\
&+ g(1-\delta_{ij}) \frac{\partial^2 V_2(|\bm{r}_i-\bm{r}_j|)}{\partial x_i^\mu \partial x_i^\nu},
\label{eq:Hessian}
\end{align}
where $x_i^\mu, x_i^\nu = x_i,y_i,z_i$ and $\delta_{ij}$ is the Kronecker delta.
The minimum eigenvalue $\lambda$ of Eq.(\ref{eq:Hessian}) is obtained which is classified by the type of 
lattice displacement. 
When the two-body potential is absent at $g = 0$ we have four-fold degenerate eigenvalues $\lambda=\lambda_2$ of the one-body potential $V_1$, which are the three 2-$in$-2-$out$ modes and one all-$in$-all-$out$ mode. 
A finite $g$ or namely $V_2$-term lifts the degeneracy and the 2-$in$-2-$out$ becomes the softest mode as shown in Fig.~\ref{tetra} (b). Here, we applied the standard form of elastic energy, $V_2(\bm r)=(|\bm r|-a_2)^2/2$, without the loss of generality. In this way, the ice-type lattice distortion is selected. 

\section{Energy splitting by Jahn-Teller effect} 
\label{sec_jti}
\begin{figure*}[t]
\includegraphics[clip,width=160mm]{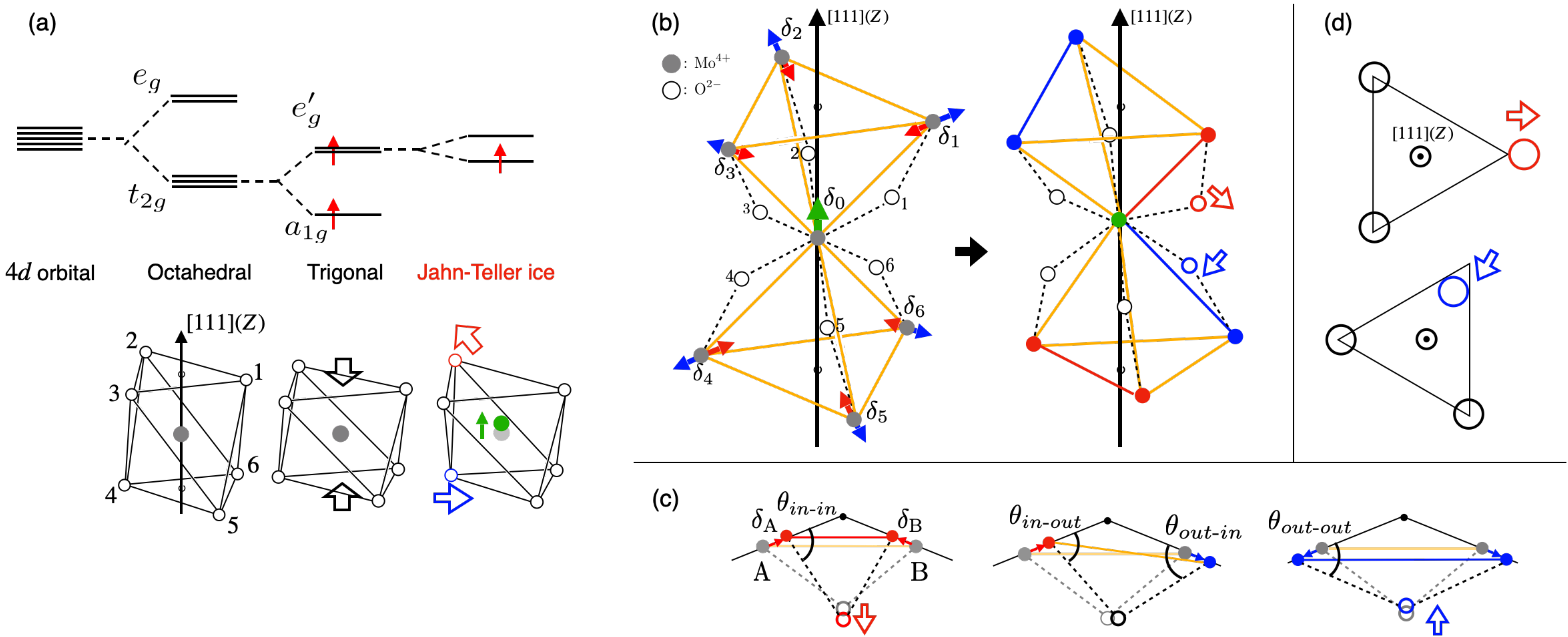}
\caption{
(a) Lifting of the $e_g'$ orbital degeneracy.
(b) Local structure before (left) and after 2-$in$-2-$out$ distortion (right).
The red and blue lines represent the $in$-$in$ and $out$-$out$ bonds, respectively.
The oxygen ion between $in$-$in$ bonds is displaced away from the tetrahedron while the one between $out$-$out$ bonds gets closer. 
(c) The oxygen displacements of $in$-$in$, $in$-$out$ and $out$-$out$ bonds. 
Black dots represent the center of the Mo$_4$ tetrahedron.
(d) The oxygen displacements viewed from directly above.
}
\label{crystalfield}
\end{figure*}
The trigonal crystal field splits the $t_{2g}$ orbitals into the $a_{1g}$ and the doubly-degenerate $e_g'$ states, as shown in Fig.~\ref{crystalfield}(a). 
One of the two $d$ electrons on the Mo$^{4+}$ is accommodated in the $a_{1g}$ orbital and the other in one of the doubly-degenerate $e_g'$ orbitals, taking the high spin state based on Hund's rule.
The doubly-degenerate $e_g'$ orbitals are JT active.
\par
The $in$-$out$ displacement of a single Mo$^{4+}$ ion alone is not enough to lift the degeneracy of the $e_g'$-orbitals. 
In fact, when the six O$^{2-}$ ions surrounding the Mo$^{4+}$ ion stay in the regular position, 
the displacement of Mo$^{4+}$ along the $Z$-axis does not break the trigonal symmetry. 
The symmetry is broken when the O$^{2-}$ ion between two neighboring Mo$^{4+}$ ions 
moves together as shown in Fig.~\ref{crystalfield}(b). 
The X-ray measurements show that the fluctuation of the Mo-O distance is smaller than that of the Mo-Mo distance by an order of magnitude \cite{booth2000local}.
This means that Mo$^{4+}$ and O$^{2-}$ ions move coherently overall and the Mo-O distances remain unchanged. 
To be precise, the distortion of two adjacent Mo$^{4+}$ ions is coupled to the displacement of the O$^{2-}$ ion in between them; 
if the two Mo$^{4+}$ ions move both inside the tetrahedra ($in$-$in$), the O$^{2-}$ moves away from the Mo-Mo bond, 
and if Mo$^{4+}$ ions take the $out$-$out$ configuration the O$^{2-}$ ion moves closer to the Mo-Mo bond 
as shown in Fig.~\ref{crystalfield}(c).
Accordingly, the relative positions of O$^{2-}$ ions given in Eqs. (1-3) are modified. 
Note that the longitude $\phi$'s are invariant for the $in$-$out$ displacements shown in Fig. \ref{crystalfield} (d). 
\par
Now, we consider a neighboring A-B pair of Mo$^{4+}$ ions, and the O$^{2-}$ ion between them. We define the displacement of the Mo$^{4+}$ ions as $\delta_{\rm A}, \delta_{\rm B}$ whose sign corresponds to $in~(+)$ or $out~(-)$ as shown in Fig.~\ref{crystalfield} (c). 
The relative angle of O$^{2-}$ from the $i$-th Mo$^{4+}$ ion is obtained up to the first order of $(\delta/a)$ as 
\beq
\theta_{\delta_{\rm A}\text{-}\delta_{\rm B}} = \theta_0 + \Gamma_{\rm A}\frac{\delta_{\rm A}}{a}+ \Gamma_{\rm B}\frac{\delta_{\rm B}}{a},
\label{theta_in_out}
\eeq
where
\begin{eqnarray}
&&\Gamma_{\rm A} = \frac{\sin \theta_0 + \cos \theta_0/2\sqrt{2}}{(\sin \theta_0 - \cos \theta_0 / \sqrt{2})(\sin \theta_0 + \sqrt{2}\cos \theta_0)},
\\
&&\Gamma_{\rm B} = \frac{3\cos \theta_0/2\sqrt{2}}{(\sin \theta_0 - \cos \theta_0 / \sqrt{2})(\sin \theta_0 + \sqrt{2}\cos \theta_0)}.
\end{eqnarray}
Then, the $\delta$-dependence of the $e_g'$ energy levels is evaluated as follows. 
Using the expansion with spherical harmonic functions $Y_{km} (\theta,\phi)$, the Coulomb potential $v(\bm{r})$ from the crystal field for $d$-electrons is given by,
\begin{align}
\nonumber
v_{\rm cry}(\bm{r}) = A_{00} &+ \sum_{m=-2}^2A_{2m} r^2 C_m^{(2)}(\theta,\phi) \\
&+ \sum_{m=-4}^4A_{4m} r^4 C_m^{(4)}(\theta,\phi)
\label{expand_pote}
\end{align}
with
\begin{align}
A_{km} &= \sqrt{\frac{4\pi}{2k + 1}} \frac{Ze^2}{a^{k+1}} \sum_{i=1}^{6} Y_{km}^* (\theta_i,\phi_i), \label{coefficient} \\
C_m^{(k)}(\theta,\phi) &= \sqrt{\frac{4\pi}{2k + 1}} Y_{km} (\theta,\phi).
\end{align}
Here, $Z=2$ is an ionic charge of O$^{2-}$ and $e$ is the elementary charge.
The coefficient $A_{km}$ depends on the relative positions of the ligands from the Mo$^{4+}$ ion. 
In the case of the trigonal crystal field, we find that $A_{km}$ is non-zero for $m\neq 0, \pm3$ because of the 3-fold rotational symmetry,  and $A_{km} = (-1)^mA_{k-m}$ due to the reflection symmetry with respect to the $X$-$Z$ plane.
Hence, the crystal potential of the trigonal crystal field is obtained as,
\begin{align}
\nonumber
v_{\rm tri}(\bm{r}) &= A_{00}^{\rm tri} +  A_{20}^{\rm tri} r^2 C_0^{(2)}(\theta,\varphi)+ A_{40}^{\rm tri} r^4 C_0^{(4)}(\theta,\varphi) \\
& + A_{43}^{\rm tri} r^4 \qty[C_3^{(4)}(\theta,\varphi) - C_{-3}^{(4)}(\theta,\varphi)],
\label{tri_potential}
\end{align}
where $A_{km}^{\rm tri}$ represents the coefficient given in \eq{coefficient} for the trigonal crystal field.
Hereafter, we suppose that the electron is localized enough, 
and consider the crystal field potential up to the second order with respect to $(r/a)$ 
discarding higher order terms.
The $a_{1g}$ and $e_g'$ orbitals which are the eigenstates of the potential are
\begin{align}
\ket{a_{1g}}&= R_{42}(r)Y_{20}(\theta,\phi), \\
\ket{e_{g\pm}'}&= \mp \frac{1}{\sqrt{3}}R_{42}(r)(\sqrt{2}Y_{2\mp2}(\theta,\phi) \pm Y_{2\pm1}(\theta,\phi)),
\end{align}
where $R_{42}(r)$ is the radial part of the wave vector.
The orbital energies of $a_{1g}$ and $e_{g}'$ are obtained as
\begin{align}
E_{a_{1g}} &= \bra{a_{1g}}v_{\rm tri}(\bm{r}) \ket{a_{1g}} = \frac{2}{7} A_{20}^{\rm tri}\ol{r^2}, \\
E_{e_g'} &= \bra{e_{g\pm}'}v_{\rm tri}(\bm{r}) \ket{e_{g\pm}'} = -\frac{1}{7} A_{20}^{\rm tri}\ol{r^2},
\end{align}
where
\begin{align}
\overline{r^2} = \int_0^\infty r^2 |R_{42}(r)|^2r^2dr = 504 \qty( \frac{a_{\rm B}}{Z_{\rm Mo}})^2.
\end{align}
Here, $a_{\rm B} = 0.529 (\AA)$ is the Bohr radius and $Z_{\rm Mo}=42$ is the atomic number of molybdenum.
The coefficient
\begin{align}
  A_{20}^{\rm tri} = 3Ze^2(3\cos^2 \theta_0 -1)/a^3
\end{align}
  takes a negative value if $\theta_0 > \theta_0^{\rm oct} = \cos^{-1}(1/\sqrt{3})$.
Note that the off-diagonal components are zero since all bases are orthogonal to each other, i.e. $\bra{e_{g\pm}'}v_{\rm tri}(\bm{r}) \ket{e_{g\mp}'} = \bra{e_{g\pm}'}v_{\rm tri}(\bm{r}) \ket{e_{g\pm}'} = 0$. 
\par
We now treat the effect of lattice distortion $\delta/a\ll 1$, 
which modifies the angle in Eq.(\ref{theta_in_out}) as perturbation to the trigonal crystal field. 
The degeneracy of the $e_g'$ orbital is lifted by the perturbation if the off-diagonal component $\bra{e_{g\pm}'}v_{\rm tri}(\bm{r}) \ket{e_{g\mp}'}$ takes a nonzero value, which is given by
\beq
 \bra{e_{g+}'}v_{\rm ice}(\bm{r}) \ket{e_{g-}'} = \frac{\sqrt{6}}{21}(2\sqrt{2}A_{2-1}^{\rm ice}-A_{22}^{\rm ice})\overline{r^2},
 \label{ene_ice1}
 \eeq
where $A_{km}^{\rm ice}$ represents the coefficient given in \eq{coefficient} for the perturbed trigonal crystal field. 
From the eigen equation, 
\beq
\mqty|E_{e_g'} - \lambda & \bra{e_{g+}'}v_{\rm ice} \ket{e_{g-}'}\\ \bra{e_{g+}'}v_{\rm ice} \ket{e_{g-}'}^* & E_{e_g'} - \lambda| = 0,
\eeq
we obtain 
\beq
\lambda = E_{e_g'} \pm \qty|\bra{e_{g+}'}v_{\rm ice}(\bm{r}) \ket{e_{g-}'}|, 
\eeq
and the off-diagonal component determines the degree of the splitting of energy levels.
\par
We consider a situation where the central Mo$^{4+}$ ion moves into the upper tetrahedron, 
i.e. the displacement $\delta_0 > 0$ (See Fig. \ref{crystalfield} (b)). 
Using the variable $\delta_j$ on the nearest-neighboring site $(j = 1,2,...,6)$ and \eq{theta_in_out}, 
the angles $\theta_j$ of six Mo-O bonds about the $Z$-axis can be written as
\beq
\theta_j = 
\begin{cases}
\displaystyle
\theta_0 + \qty(\Gamma_{\rm A}\frac{\delta_0}{a} +\Gamma_{\rm B}\frac{\delta_j}{a})&j=1,2,3 \\
\displaystyle
\pi - \theta_0 + \qty(\Gamma_{\rm A}\frac{\delta_0}{a} -\Gamma_{\rm B}\frac{\delta_j}{a})&j=4,5,6. 
\label{eq:thetamod}
\end{cases}
\eeq
Substituting Eq.(\ref{eq:thetamod}) in \eq{coefficient} and using \eq{ene_ice1}, 
we find that the energy splitting depends on $\qty{\delta_{j}}$ up to the first order of $(\delta/a)$ as
\beq
\qty| \bra{e_{g+}'}v_{\rm ice}(\bm{r}) \ket{e_{g-}'}| = \frac{\eta}{a}\qty|P(\qty{\delta_{j\in \partial 0}})| ,
\label{eg_pdelta}
\eeq
where 
\beq
\eta = \frac{1}{14}(-4\sqrt{2}\cos2\theta_0 + \sin 2 \theta_0)\Gamma_{\rm B}\frac{\ol{r^2}}{a^2}\frac{Ze^2}{a},
\eeq
and
\beq
P(\qty{\delta_{j\in \partial 0}}) 
= (\delta_1 + \delta_4) + (\delta_2+ \delta_5)e^{\frac{2\pi }{3}i} + (\delta_3+ \delta_6)e^{\frac{4\pi}{3}i}.
\label{p_delta}
\eeq
Here, $\delta_{j \in \partial i}$ denotes the indices of the six surrounding Mo$^{4+}$ ions 
centered by the $i$-th Mo$^{4+}$ ion. 
Interestingly, the energy splitting does not depend on the displacement of the central Mo$^{4+}$ ion, $\delta_0$, because the contribution is canceled out between the upper and lower tetrahedra.
To briefly summarize, we obtained the orbital-energy splitting of a Mo$^{4+}$ ion 
in the linearly combined form of the displacements $\{ \delta_j \}$ of the six adjacent Mo$^{4+}$ ions. 

\begin{table}[t]
  \begin{tabular}{|c|c|c|c|c|} \hline
   & ~CF symmetry~  & ~$e_{\rm JT}^{\rm min}$~ & ~$(\sigma_1,\sigma_2,\sigma_3;\sigma_4,\sigma_5,\sigma_6)$~ & ~$\Phi$~\\ \hline
   (\gone) & T$_{\rm u}$, T$_{\rm l}$, R & 0 & $(+,+,+;\mp,\mp,\mp)$   & $0, \pm6$\\ \hline
   (\gtwo) & T$_{\rm u}$, R  & $-4\epsilon$ & $(+,+,+;\mp,\pm,\pm)$ & $\pm4, \pm2$ \\ \hline
   (\gthree) & $-$  & $-4\epsilon$   & $(+,-,-;-,+,-)$ & $\pm2$ \\ \hline  
   (\gfour) & R & $-16\epsilon$ & $(+,-,-;+,-,-)$ & $\pm2$  \\ \hline
   (\gfive) & $-$& $-12\epsilon$ & $(+,-,-;+,-,+)$  & $0$ \\ \hline
   (\gsix) & R  & 0 & $(+,-,-;-,+,+)$ & $0$ \\ \hline
  \end{tabular}
  \caption{JT energy $e_{\rm JT}^{\rm min}$ of a single Mo$^{4+}$ ion surrounded by six Mo$^{4+}$ 
    with displacements $\{\sigma_{j\in \partial 0}\}$, $i=1\sim 6$. 
    T$_{\rm u}$, T$_{\rm l}$ and R represent the three-fold rotational symmetry of the upper tetrahedron, 
    that of the lower tetrahedron, and the reflection symmetry, respectively. 
    The $in/out$ displacement of the center Mo$^{4+}$ ion ($\sigma_0 = \pm1$) 
    does not change the crystal field or $e_{\rm JT}^{\rm min}$. 
    The symmetry operations on $\{\sigma_j\}$ that do not change the crystal field of all patterns are 
    tuning over all displacements $\{\sigma_j\} \to \{-\sigma_j\}$ and/or exchanging the configurations of 
    upper and lower triangles as $(\sigma_1,\sigma_2,\sigma_3) \leftrightarrow (\sigma_4,\sigma_5,\sigma_6)$. 
    Right panel shows the example of lattice displacements and the surrounding six O$^{2-}$ ions for case (\gtwo). 
    The total flux that comes in and out of the pair of tetrahedra classifies the energy of cases (\gfour) and (\gsix). 
    For the lattice distortion patterns listed here, see Fig.~\ref{jt_ice}. 
    }
   \label{table1}
\end{table}
\begin{figure*}[t]
\includegraphics[clip,width=150mm]{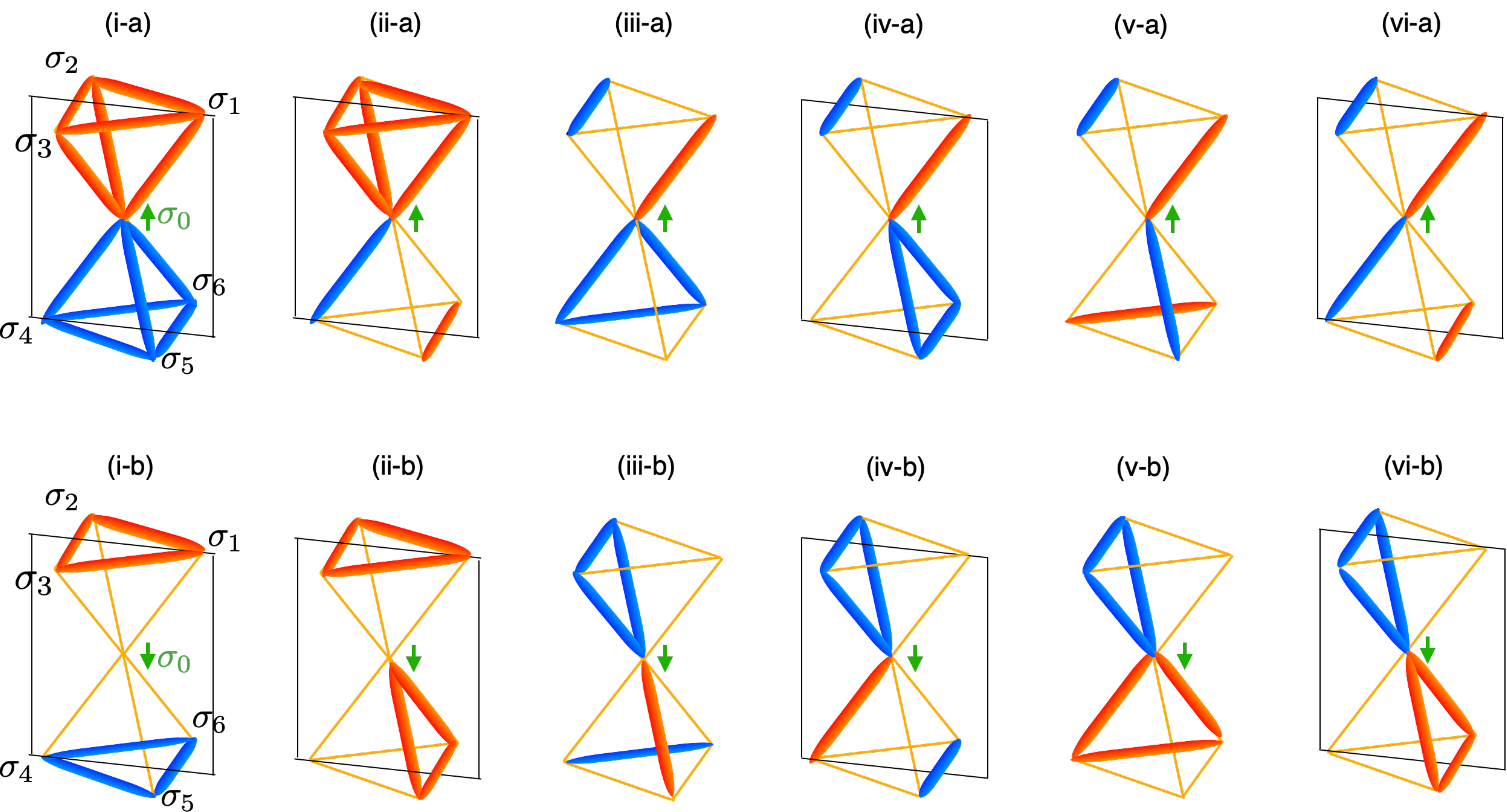}
\caption{
Displacements of Mo$^{4+}$ ions corresponding to (\gone)-(\gsix) given in TABLE I. 
For each pattern, the two cases where the center Mo$^{4+}$ ion moves to
the upper tetrahedron (a) [top row] and moves to the lower tetrahedron (b) [bottom row] are shown.
The local energy associated with the center Mo$^{4+}$ ion is the same for (a) and (b) (see text). 
Red and blue bonds represent $in$-$in$ and $out$-$out$ bonds, respectively.
}
\label{jt_ice}
\end{figure*}
\section{Microscopic Hamiltonian of the Jahn-Teller ice } \label{sec_simplify}
The JT energy is the sum of the elastic energy and the orbital-energy splitting. 
For the latter, we have so far focused on ``local" orbital energy gain on a single Mo$^{4+}$ ion 
with displacement $\delta_0$, given as Eqs.(\ref{eg_pdelta}) and (\ref{p_delta}). 
By summing up these local contributions, 
the JT Hamiltonian of the whole system $(i=1,2,...,N)$ is given as
\beq
H = \frac{\lambda}{2}  \sum_{i=1}^N \delta_i^2 - \frac{\eta}{a}\sum_{i=1}^N \qty|P(\qty{\delta_{j \in \partial i}})|. 
\label{hbare0}
\eeq
For simplicity, we consider only the softest $in/out$ displacement of the Mo$^{4+}$ ions, 
which safely allows us to abbreviate the $V_{2}$ term in \eq{vib_ene}. 
The parameter $\lambda$ is the elastic energy scale corresponding to $\lambda_{2}$ in \eq{ellips}. 
\par
The essential energetics of Eq. (\ref{hbare0}) can be described more simply by 
introducing a set of Ising variables $\{\sigma_j\}$ with $\sigma_j=\pm 1$  
that represents the $in$ and $out$ lattice displacements. 
Here, we replace $\delta_{j \in \partial i} \rightarrow \bar{\delta_i} \sigma_{j \in \partial i}$ 
by approximating $\bar{\delta}_i>0$ as the ``locally averaged" amplitude of the lattice displacements 
around the $i$-th  Mo$^{4+}$ ion. 
The Hamiltonian is rewritten as $H = \sum_{i=1}^N e_{\rm JT}^{(i)}$ with 
\beq
e_{\rm JT}^{(i)} =
\frac{\lambda}{2} \bar{\delta_i}^2 -  \bar{\delta_i} \frac{\eta}{a} \qty|P^\prime(\qty{\sigma_{j \in \partial i}})|, 
\label{ene_l}
\eeq
Here, the first term of \eq{ene_l} representing the mean elastic energy 
can be intuitively regarded as that of the Einstein model taking the oscillation unit as a pair of tetrahedra sharing the 
$i$-th Mo$^{4+}$ ion. 
The term for the orbital energy splitting in \eq{p_delta} is rewritten as
\beq
P^\prime(\qty{\sigma_{j\in \partial 0}}) = (\sigma_1 + \sigma_4) + (\sigma_2+ \sigma_5)e^{\frac{2\pi }{3}i} 
+ (\sigma_3+ \sigma_6)e^{\frac{4\pi}{3}i}. 
\label{p_sigma}
\eeq
\eq{ene_l} takes the minimum value, 
\beq
e_{\rm JT}^{(i){\rm min}} = - \epsilon|P'(\qty{\sigma_{j \in \partial i}})|^2,\;\; \epsilon = \frac{\eta^2}{2\lambda}
\label{epsilon}
\eeq
when 
\beq
\frac{\bar{\delta}^*_i}{a} = \frac{\eta|P'(\qty{\sigma_{j \in \partial i}})|}{\lambda}.
\label{deltastar}
\eeq
The value of $\bar{\delta_i}^*$ is locally determined by $|P'(\qty{\sigma_{j \in \partial i}})|$ and may depend on $i$. 
This approximates well the situation where $|\delta_i|$ can vary depending on $i$ 
when considering the Hamiltonian defined as a summation of \eq{p_delta} of all ions. 
However, the only important point here is that the elastic energy increases in square and the orbital splitting energy 
increases linearly in $\delta_i$, which guarantees that there is a finite value $|\delta_i| >0$ that optimize 
the total energy. It is natural to expect that such $|\delta_i|$ does not vary much from ion to ion. 
We confirmed numerically in a small size cluster that the results basically remain 
qualitatively unchanged even if we approximate $|\delta_i|$ to be $i$-independent. 
\par
Since $|P'(\qty{\sigma_{j \in \partial i}})|$ and $|P'(\qty{\sigma_{j \in \partial k}})|$ share some $\sigma_j$'s and thus are correlated, the summation of the local minimums of $-|P'(\qty{\sigma_{j \in \partial i}})|$ are not necessarily the global minimum. 
To see this in more detail, we focus on the local JT energy gain $e_{\rm JT}^{(0){\rm min}}$ around the 0-th Mo$^{4+}$ ion in \eq{epsilon}. We stress here that it does not depend on its own displacement, $\sigma_0$, but on other Mo$^{4+}$ ion's $\sigma_j$. Table \ref{table1} displays the value of the JT energy $e_{\rm JT}^{(0){\rm min}}$ for all distortion patterns $(\gone)$-$(\gsix)$(see  Fig.~\ref{jt_ice} for the corresponding images). Here, these patterns are classified according to the symmetry of the crystal field. From \eq{p_sigma} it is obvious that the three-fold rotational symmetry of the upper and lower triangles will erase the terms and suppress the JT energy gain, which is reflected in cases $(\gone)$ and $(\gtwo)$. The reflection symmetry in $(\gfour)$ and $(\gsix)$ works in different manners; for the case $(\gfour)$ if the pair ($\sigma_2,\sigma_3$) on the upper triangle are the same and if they are also the same as ($\sigma_4,\sigma_6$), they will cooperatively increase $|P'|$, whereas for the case $(\gsix)$, the latter has a different sign from the former and suppresses $|P'|$. 
For cases $(\gthree)$ and $(\gfive)$ the above mentioned symmetries are absent, while the latter has slightly higher symmetry between the upper and lower triangles; in case (\gfive), rotating the upper triangle by $2\pi/3$ and by turning over $(\sigma_1,\sigma_2,\sigma_3)$, it matches $(\sigma_4,\sigma_5,\sigma_6)$, 
which means that apart from the small cancellation of phase factors, $|P'|$ is relatively large. 
Case $(\gthree)$ has much lower symmetry and has smaller $|P'|$. 
\par
For later convenience on connecting the $in$-$in$ and $out$-$out$ bonds of more numbers of tetrahedra, 
we also classify cases $(\gone)$-$(\gsix)$ by the total flux defined as 
\beq
\Phi=\sum_{i=1}^6\sigma_i. 
\eeq
If $\Phi >0$ or $<0$ the flux flows in or out of a pair of tetrahedra. 
The cases $\Psi=0$ show some sort of symmetry between upper and lower triangles, 
which are fulfilled in cases (\gone) and (\gsix). 
\par
An important difference from the ice rule is that the $in$-$out$ configurations do not necessarily show 
one-to-one correspondence with the JT energy. 
For example, as shown in Figs.~\ref{jt_ice} $(\gone\text{-a},\gone\text{-b})$, depending on whether the 
center Mo$^{4+}$ moves upward or downward, the $(\gone\text{-a})$ 4-$in$/4-$out$ and $(\gone\text{-b})$ 3/1-$in$-1/3-$out$ 
are realized but its own $|P'|$ at the center Mo$^{4+}$ remains the same. 
Notice, however, it {\it does} change the $|P'|$'s of the six surrounding Mo$^{4+}$'s. 
\par
Indeed, the total JT energy depends sensitively on the combinations of these 12 patterns over the whole lattice. The locally lowest energy  $-16\epsilon$ is realized in case $(\gfour)$ with $\{\sigma_{j \in \partial 0}\} = (\sigma_1,\sigma_2,\sigma_3;\sigma_4,\sigma_5,\sigma_6) = (+,-,-;+,-,-)$. However, this distortion gives the 2-$in$-2-$out$ pattern for upper/lower tetrahedra, but the lower/upper tetrahedra have 1-$in$-3-$out$, which keeps the 3-fold rotational symmetry about another $Z$-axis defined along the connection to the other tetrahedron. This will raise the JT energy of the neighboring Mo$^{4+}$ ion on that axis. 
\par
On the other hand, in case $(\gfive\text{-a})$ with $\{\sigma_{j \in \partial 0}\} = (+,-,-;+,-,+)$,
the distortion consists of only 2-$in$-2-$out$ types. 
The energy becomes $-12\epsilon$ and JT energies of the neighboring Mo$^{4+}$ ions can afford $-12\epsilon$ as well. 
It is expected that the bulk JT energy resultantly becomes the lowest when all the tetrahedra 
take the 2-$in$-2-$out$ structure, i.e. the ground state. 
Notice that although Fig.~\ref{jt_ice}$(\gfive\text{-b})$ is the 3-$in$-1-$out$/1-$in$-3-$out$ 
it has the same $-12\epsilon$, whereas these two differ in the JT energy about the six $\{\sigma_j\}$ ions, 
and hence the non 2-$in$-2-$out$ structure pattern tends to raise the total JT energy. 
\par
In this way, although a given set of $\{\sigma_j\}$ $j=1\sim 6$ will determine 
the local JT energy based on Eq.(\ref{p_sigma}), the $|P'|$'s belonging to the neighboring tetrahedra are strongly correlated. 
To provide better physical insights into this correlation effect, 
we construct an effective Hamiltonian of the whole system 
which faithfully reproduces the JT energy in Table \ref{table1} for all local variables: 
\beq
H_\sigma  = \epsilon \bigg( 2\sum_{\la i,j \ra}\sigma_i \sigma_j + \sum_{\langle \! \langle  i,j \rangle \! \rangle}\sigma_i \sigma_j - 2 \sum_{\langle \! \langle \! \langle i,j \rangle \! \rangle \! \rangle}\sigma_i \sigma_j\bigg) -6\epsilon N,
\label{ene_all}
\eeq
where $\la i,j \ra,~\langle \! \langle  i,j \rangle \! \rangle$ and $\langle \! \langle \! \langle i,j \rangle \! \rangle \! \rangle$ denote the summations over the nearest-neighbor (NN) Mo$^{4+}$-displacement pairs, 
the second-NN pairs and the third-NN pairs, respectively. 
For example, the second-NN pairs are $(\sigma_1,\sigma_6)$, 
and the third-NN pairs are $(\sigma_1,\sigma_4)$ and $(\sigma_1,\sigma_5)$ in Fig.~\ref{jt_ice}. 
In the pyrochlore lattice, we have two species of tetrahedra with dark and light colors 
in Fig.~\ref{fig_supercooling}(a), 
where we take $\sigma_i=+1/-1$ when the Mo$^{4+}$ ions move into/out the dark tetrahedra. 
\par
\begin{figure}[t]
\includegraphics[clip,width=85mm]{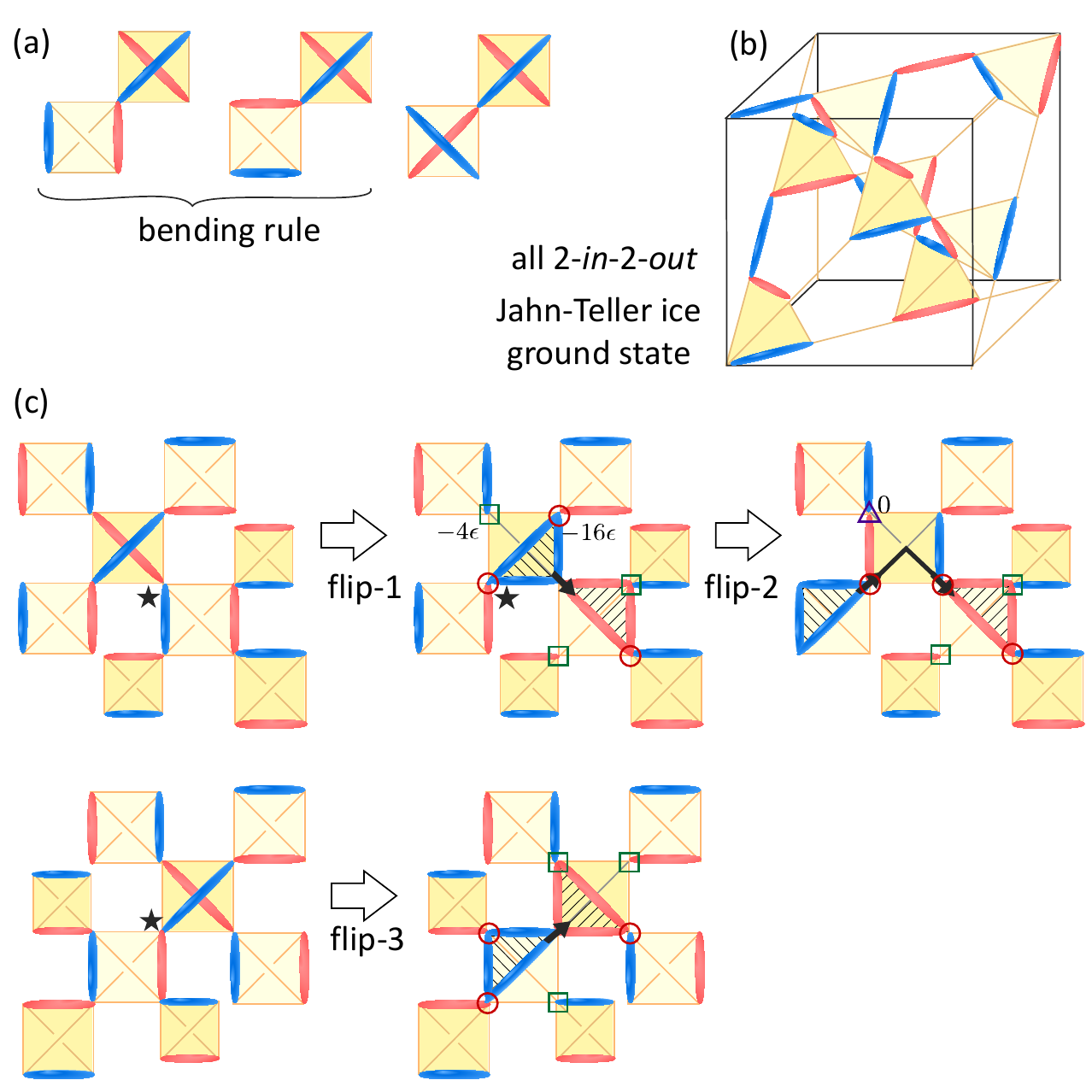}
\caption{(a) Three choices of bond connections between two tetrahedra following the ice rule 
when the upper-right one is given. 
The first two patterns fulfill the bending ice rule and are allowed in the Jahn-Teller ice. 
Blue and red bonds represent $in$-$in$ and $out$-$out$ bonds, respectively. 
(b) Example of the configuration of bonds following the bending ice rule, 
which forms the ground state. 
(c) Examples of flipping $\sigma_i$, namely converting $in/out$ to $out/in$ of an ion marked with a star 
from the Jahn-Teller ice. 
Circles/squares/triangles are the $-16\epsilon$,$-4\epsilon$,$0$ ions, respectively. 
The $-12\epsilon$ ions forming the ground state are not marked. 
We call the shaded triangles formed by 3-$in$ or 3-$out$ as monopoles of spin ice. 
}
\label{jti_gs_scl}
\end{figure}

The first term of \eq{ene_all}, i.e., the NN interaction, is the same as the spin ice model that remains disordered down to zero temperature whose ground state is macroscopically degenerate. In general, adding second-NN and third-NN interactions lifts this ground state degeneracy except for the specific parameter sets \cite{ikeda2008ordering, rau2016spin, udagawa2016out}. 
In our \eq{ene_all}, these longer-range interactions are as strong as the NN interaction so that they cannot be regarded merely as perturbations; the important consequence is a {\it bending ice rule}. Under the standard ice rule, all the tetrahedra have one red bond and one blue bond. We show in Fig.~\ref{jti_gs_scl}(a) two connected tetrahedra viewed from the top. If we place a certain pair of red-blue bonds at the upper-right tetrahedron, the lower-left tetrahedron can take three different patterns, where the bonds with different colors are always connected one to one. However, as we saw for case (\gfive) in Table I, our JT ice rule prohibits the two connected bonds from aligning in the same direction. We call this a bending ice rule, that excludes the third pattern in Fig.~\ref{jti_gs_scl}(a). Figure~\ref{jti_gs_scl}(b) shows one of the ground states that fulfills the bending ice rule: the red and blue bonds alternatively align and are bent at all vertices. Here, a repeated structure in a unit cell period is shown, and there are twelve such regular states. Although one cannot exclude the possibility that there are several nonperiodic patterns of the same energy, we confirmed that the bending rule throughout the crystal will significantly reduce the number of configurations of the ground state, which will be discussed in Appendix \ref{sec-icerules}. 
\par
Still, one may suspect other possibilities that there can be a non-2-$in$-2-$out$ structure with $-12\epsilon$ that join the ground state, or the three $-12\epsilon$ tetrahedra can be replaced by two $-16\epsilon$ and one $-4\epsilon$ which we call a ``trimer". 
To understand the difficulty of lowering the energy by including such irregular structures beyond the bending ice rule, 
we show in Fig.~\ref{jti_gs_scl}(c) some examples starting from the pattern allowed by the bending ice rule. 
In the flip-1 process, the ion marked with a star moves out, and a trimer ($-16\epsilon, -16\epsilon, -4\epsilon$) is created. 
However, this process will influence the energy of the other three ions on the lower right tetrahedron, 
creating one $-16\epsilon$ and two $-4\epsilon$ which are higher in energy. 
\par
Unlike the standard ice rule, not only the ions on the two tetrahedra sharing a star-site but 
those belonging to the other six tetrahedra around them are influenced by a single flip. 
The created 1-$in$-3-$out$ and 3-$in$-1-$out$ are called monopoles following the convention of spin ice, 
although these two are not free and thus are not true monopoles. 
When we perform flip-2, the former monopole moves to the lower-left, and the center tetrahedron recovers 2-$in$-2-$out$. 
Here is another difference from the freely moving monopoles in the spin ice: 
the bending rule is broken and the energy increases significantly. 
Flip-3 is another example of creating a pair of monopoles, which shows similar energetics. 
These examples imply that creating irregular structures will raise energy in each process. 
Although there might be some chance of creating a locally stable trimer $(-16\epsilon,-16\epsilon,-4\epsilon)$, 
surrounded by $-12\epsilon$'s 
when approaching from the high-temperature random configuration, once such local structure is created 
it is difficult to lower the energy except by overcoming a high energy barrier over several flips. 
\par
In the next section, 
we find rather unexpectedly that the liquid state survives down to the lowest temperatures 
as a meta-stable liquid state, which is the consequence of the above-mentioned energetics. 

\section{Monte Carlo simulation}
\label{sec-MC}

To investigate the nature of the JT ice model at finite temperatures,
we perform a Monte Carlo simulation based on the effective Hamiltonian \eq{ene_all}
at finite temperatures.
We consider a periodic system of cubic geometry consisting of $L^3$ unit cells 
with a total number of lattice sites $N=16L^3$. 
At each Monte Carlo step (MCS), we sequentially perform a single-lattice-flip Metropolis update for all lattice sites.
In the following, we discuss the thermodynamic, structural, and dynamic properties of the model. 

\subsection{Thermodynamic properties}
In measuring the thermodynamic properties, 
we perform both the cooling and heating simulations. 
In cooling, fully equilibrated initial configurations are prepared at a high enough temperature $k_{\rm B}T/\epsilon = 6.00$ 
and the temperature is lowered by a small step $\Delta (k_{\rm B}T/\epsilon) = 0.01$, 
where we take $\tau_{\rm cool}=1.0\times 10^6$ MCS for equilibration and the same $1.0\times 10^6$ MCS 
for taking a thermal average $\expval{\cdots}$ at each temperature. 
The cooling rate dependencies are separately examined for $\tau_{\rm cool}=1.0\times 10^1$ to $10^5$. 
We perform 10 independent runs for the system size $L=6,8,9,10,12$ and evaluate 
the statistical errors of the observables. 
For heating ($L=9$), we construct ground state configurations explicitly as the initial configurations, 
and take $\Delta (k_{\rm B}T/\epsilon) =0.01$ with $1.0\times 10^6$ MCS for both equilibration and thermal average. 

\begin{figure}[t]
\includegraphics[width=65mm]{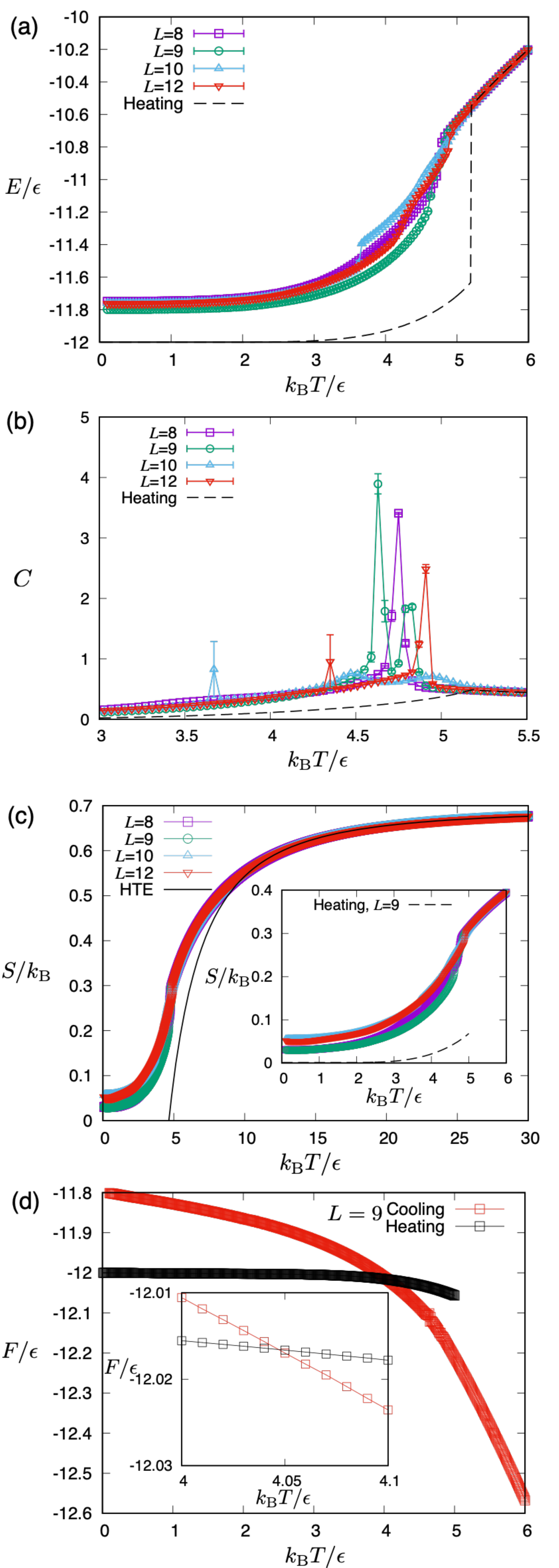}
\caption{
  Temperature dependence of thermodynamic quantities 
  obtained by the Monte Carlo simulations for different system sizes $L$ in the JT ice model. 
  (a) The internal energy $E$, (b) heat-capacity $C$, (c) entropy $S/k_B$ and 
  (d) free-energy $F$. The energy is shown per lattice site in unit of $\epsilon$. 
}
\label{mc_sim}
\end{figure}
\par
Figures~\ref{mc_sim}(a)-\ref{mc_sim}(d) show the thermodynamic quantities obtained by cooling and heating.
The temperature dependence of the internal energy $E/\epsilon$ 
clearly exhibits disagreement between the cooling and heating processes at $k_{\rm B}T/\epsilon \le 5.1$. 
This suggests that the system remains in a supercooled liquid state. 
In the heating protocol, the internal energy gradually increases 
up to $k_{\rm B}T/\epsilon \sim 5.2$ and exhibits a sudden jump. 
This jump may be regarded as an indication of the stability limit of the super-heated crystalline state (see Fig.~\ref{fig_supercooling}).
As we will see shortly, the equilibrium first-order transition between the liquid and crystalline state is expected 
at a lower temperature $T_{\rm c} \sim 4.05 \epsilon/k_{\rm B}$.
\par
\begin{figure}[t]
\centering
\includegraphics[clip,width=65mm]{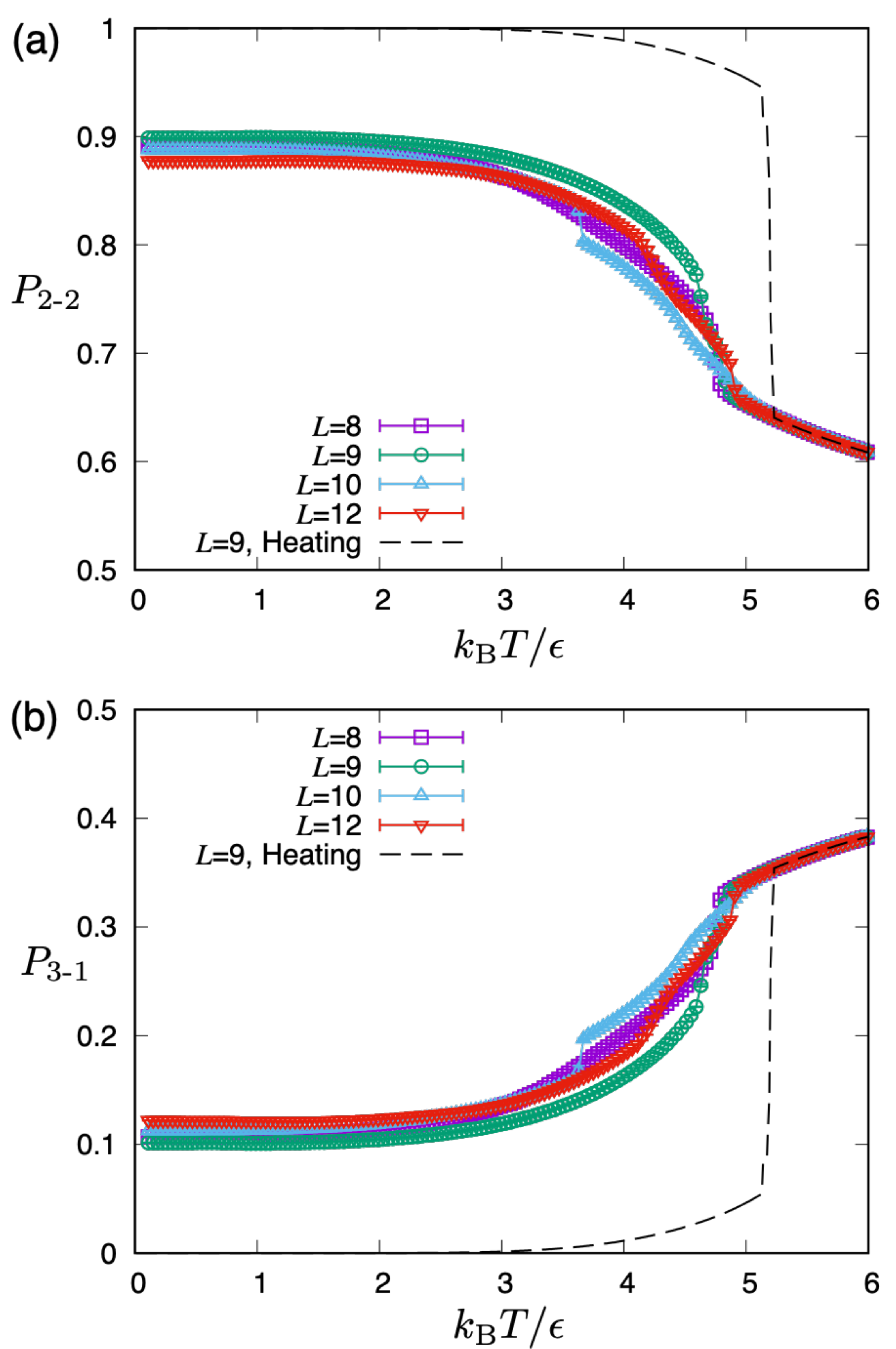}
\caption{The fraction of (a) 2-$in$-2-$out$ structure $P_{2-2}$ and 
(b) 3-$in$-1-$out$/1-$in$-3-$out$ structure $P_{3-1}$ 
of all the tetrahedra obtained by the Monte Carlo simulations for different system sizes $L$ 
in the JT ice model. 
}
\label{two_in_two_out}
\end{figure}

The internal energy obtained in the cooling protocol shows somewhat complicated size dependence. 
This is reflected in the specific heat obtained as
\beq
C/k_{\rm B} = \frac{1}{N(k_{\rm B}T)^2}(\expval{E^2} - \expval{E}^2). 
\eeq
As shown in Fig.~\ref{mc_sim}(b), there appear peak structures whose positions 
depend very sensitively on the system size $L$. 
The number of peaks also varies with $L$; a single peak in $L=8$, 
double peaks in $L=9,12$, and triple peaks in $L=10$. 
The overall profile of $C$ in the background of these peaks shows a broad maximum of around
$k_{\rm B}T/\epsilon \sim 4.5$.
\red{Since the height/width of the peaks does not systematically grow/decrease with $L$, 
we consider that they are not the indication of thermodynamic phase transitions but just the crossover where the correlation satisfying the bending ice rule grows.
Correspondingly, the fractions of the 2-$in$-2-$out$ structure and the distortion pattern $(\gfive)$ increase around the specific heat peaks (See Figs. \ref{two_in_two_out} and \ref{pattern_fraction}).
The domain growth occurs anisotropically and discretely because of the severe condition of the bending ice rule, as mentioned in Sec. \ref{sec:strct} in detail.
Therefore, }the intriguing size dependencies of the peak positions 
imply competition among several low energy configurations 
whose energies depend sensitively on the boundary condition. 
\par
We evaluate the entropy $S/k_B$ of the system associated with the crystalline 
and supercooled liquid states using the specific heat data as 
\beq
S(T) = S(T_{\rm ref})+\int_{T_{\rm ref}}^{T} \frac{dT}{T} C
\eeq
where $S(T_{\rm ref})$ is the entropy at a reference temperature $T_{\rm ref}$. 
The results are shown in Fig.~\ref{mc_sim}(c) together with 
the equilibrium entropy at high enough temperatures obtained by 
the high temperature expansion (HTE) (see Appendix \ref{sec-high-temperature-expansion} for details). 
For the crystalline state, we chose $T_{\rm ref}=0$ and $S(T_{\rm ref})=0$ 
and used the specific heat $C$ obtained by the heating protocol. 
For the supercooled liquid state we choose $S(T_{\rm ref})$ of the HTE data at $T_{\rm ref}=30 \epsilon/k_{\rm B}$ 
and use the specific heat $C$ obtained by the cooling protocol. 
The $T \to 0$ limit of the supercooled liquid branch suggests the existence of some residual entropy of $\sim 0.05k_B$. 
\par 
Using the internal energy $E$ and entropy $S$ obtained above (for $L=9$), we evaluate 
the free-energy $F=E-TS$ of the system in the crystalline and supercooled liquid
state as shown in Fig.~\ref{mc_sim}(d). 
The results indicate a first-order thermodynamic transition point 
from the liquid to crystalline state at $T_{\rm c} \sim 4.05 \epsilon/k_{\rm B}$.
\par
The cooling/heating process we observed suggests that the phase transition can be kinetically avoided easily 
even when the system is annealed extremely slowly. 
Indeed, although our cooling rate $\tau_{\rm cool}=1.0\times 10^6$ MCS is quite slow 
we do not find any sign of phase transition for all system sizes. 
The cooling rate dependence for the internal energy is examined explicitly over different 
$\tau_{\rm cool}=1.0\times 10^1$ to $1.0\times 10^{6}$ MCS
(see Fig.~\ref{mc_cooling_rate_effect}(a) in Appendix~\ref{sec-cooling-rate-effects}), 
and the data are found to gradually saturate to the behavior for the slowest cooling rate in Fig.~\ref{mc_sim}(a) 
without any hint of crystallization. 
This indicates a remarkably stable feature of our supercooled liquid state against the crystallization 
meaning that the present system can be classified as a good glass forming liquid.

\subsection{Structural properties} \label{sec:strct}
To further understand the nature of the crystalline and supercooled liquid state,
we examine the fraction of the $(4-n)$-$in$-$n$-$out$/$(n)$-$in$-$(4-n)$-$out$ structure ($n=0,1,2$) 
of the lattice distortions denoted as 
$P_{n-(4-n)}$. 
There are 16 different configurations per each tetrahedron, consisting of 
4-$in/out$, four 3-$in$-1-$out$/1-$in$-3-$out$, and six 2-$in$-2-$out$. 
\par
In the disordered high temperature limit, the 2-$in$-2-$out$ fraction approaches $P_{2-2} \rightarrow \frac{6}{16}=0.375$
 while $P_{3-1} \rightarrow \frac{8}{16}=0.5$. 
As shown in Fig.~\ref{mc_sim}(a), $P_{2-2}$ already exceeds 0.6 at $k_{\rm B}T/\epsilon \lesssim 6$ while the fraction of 
3-$in$-1-$out$/1-$in$-3-$out$ structure is $P_{3-1}\sim 0.4$ as shown in Fig.~\ref{mc_sim}(b). 
Similarly to the internal energy, the data obtained by the cooling and heating 
agree at high enough temperatures $k_{\rm B}T/\epsilon > 5.1$ but not at lower temperatures. 
In the heating protocol, for which the initial system is prepared as the ground state, 
$P_{2-2}$ remains very close to $1$ and $P_{3-1}$ to $0$ at low enough temperatures. 
\par
In the cooling protocol, 
$P_{2-2}$ shows a nearly discontinuous upturn, whose location depends on $L$ 
in accordance with the appearance of the peaks in the specific heat (See Fig.~\ref{mc_sim} (b)). 
These discontinuities are expected to disappear in the bulk limit 
since the magnitude of jump in $P_{2-2}$ decreases for larger $L$. 
Toward the low-temperature limit, $P_{2-2}$ saturates to $\sim 0.9$, which is lower than 1 of the ground state. 
The stableness of the supercooled liquid state is confirmed in the cooling rate dependence of $P_{2-2}$
(see Fig.~\ref{mc_cooling_rate_effect} (b) in Appendix~\ref{sec-cooling-rate-effects}). 
\par
Although the peaks/jumps in $C$ and $P$ are not regarded as phase transitions, 
their nonsystematic and highly sensitive $L$-dependence 
suggests that there is an underlying competition between different types of short-range orderings, 
characteristic of the frustrated systems. 
To visualize this competition, 
we calculate the static structure factor of the lattice displacement degrees of freedom 
$\sigma_i = \pm 1~(i=1,2,...,N)$ given as
\begin{align}
S_{\bm{q}} = \expval{\qty|\frac{1}{N}\sum_{i=1}^N \sigma_i e^{i \bm{r}_i \cdot \bm{q}}|^2}
\end{align}
where $\bm{r}_i$ is the position of the lattice site and 
$\bm{q} =(n_h,n_k,n_l)  \pi/L $  is the wavevector 
with integers $n_h,n_k,n_l$~$(-2L \le n_h,n_k,n_l \le 2L)$. 
Here, the unit length is taken as the side length of the cubic unit cell. 
\begin{figure*}[t]
\centering
\includegraphics[clip,width=160mm]{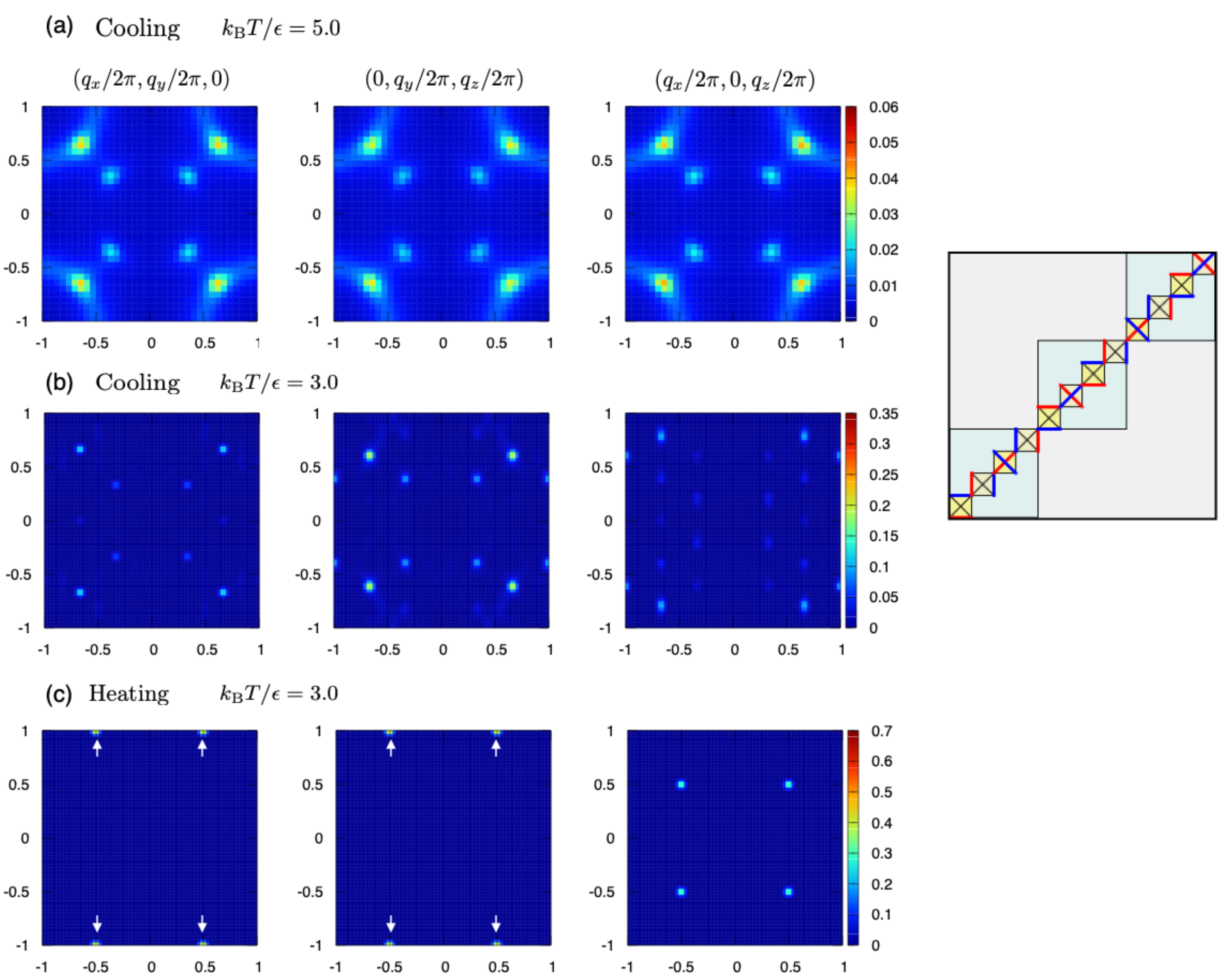}
\caption{
  Density plot of the static structure factor $S_{\bm{q}}$ of the JT ice model 
  obtained for (a) $k_{\rm B}T/\epsilon=5.0$ and (b) $k_{\rm B}T/\epsilon=3.0$ in the cooling process,
  and (c) $k_{\rm B}T/\epsilon=3.0$ in the heating process for $L=9$. 
  We plot three slices in the $\bm q$-space as 
  $\bm q=(q_x/2\pi,q_y/2\pi,0)$, $(0,q_y/2\pi,q_z/2\pi)$ and $(q_x/2\pi,0,q_z/2\pi)$. 
  We show in the right panel the three-fold periodic 2-$in$-2-$out$ structure 
  running along the [110] direction, which gives $(\pm 4/3,\pm 4/3,0)$ peak indicating the short range order. 
  The blue square indicates the size of a unit cell. 
}
\label{strct}
\end{figure*}
In Figs.~\ref{strct}(a) and ~\ref{strct}(b), we show $S_{\bm{q}}$ observed during the cooling process. 
At $k_BT/\epsilon=5.0$ the system is in the paramagnetic phase, 
where we find weak spots at $\bm{q}=(0, 2\pi/3,  2\pi/3),~(0, 4\pi/3,  4\pi/3)$ and the equivalent wavevectors. 
This peak corresponds to having a three-fold periodic structure of the 2-$in$-2-$out$ patterns. 
The example is depicted in the right panel of Fig.~\ref{strct}, 
where the bending rule shown in Fig.~\ref{jti_gs_scl}(b) is kept. 
Notice that this structure indicates the development of 
short-range order at these temperatures. 
When the system moves to the supercooled liquid region at $k_{\rm B}T/\epsilon=3.0$, 
these peaks shift slightly off these commensurate wavenumbers, and their intensity becomes larger. 
These overall peak positions agree between different $L$'s but 
their precise positions vary sensitively for different $L$ (see Appendix \ref{sec-size dependences}). 
The results mean that these incommensurate structures are stabilized from among numerous choices 
in the supercooled liquid state as the periods 
that optimally fit the particular choice of small system size by accident. 
\par
\red{In the disordered phase in a continuum or in a spin glass model on a bipartite lattice with spatially random interaction, we cannot extract any particular wavenumber, which characterizes the nature of correlation. 
However, in the spin-ice model, the disordered phase is not built on the mixture of fully different random configurations but of the ice-type configurations \cite{bramwell2020history, gardner2010magnetic}, and the structure factor exhibits a particular profile called the ``pinch point" at $(\pi/2, \pi/2,\pi/2)$.
In the same manner, the peaks of the structure factor in our calculation which we mentioned as short range order is a manifestation of the disordered state based on the bending ice rule.}
\par
For the heating process in Fig.~\ref{strct}(c) at the same $k_{\rm B}T/\epsilon=3.0$, 
the aforementioned peaks are absent, and instead, we find the peaks at $\bm q=(\pi,2\pi,0)$, 
which corresponds to the ground state configuration we chose as a regular configuration 
among several choices. 
\par
\begin{figure}[t]
\includegraphics[clip,width=65mm]{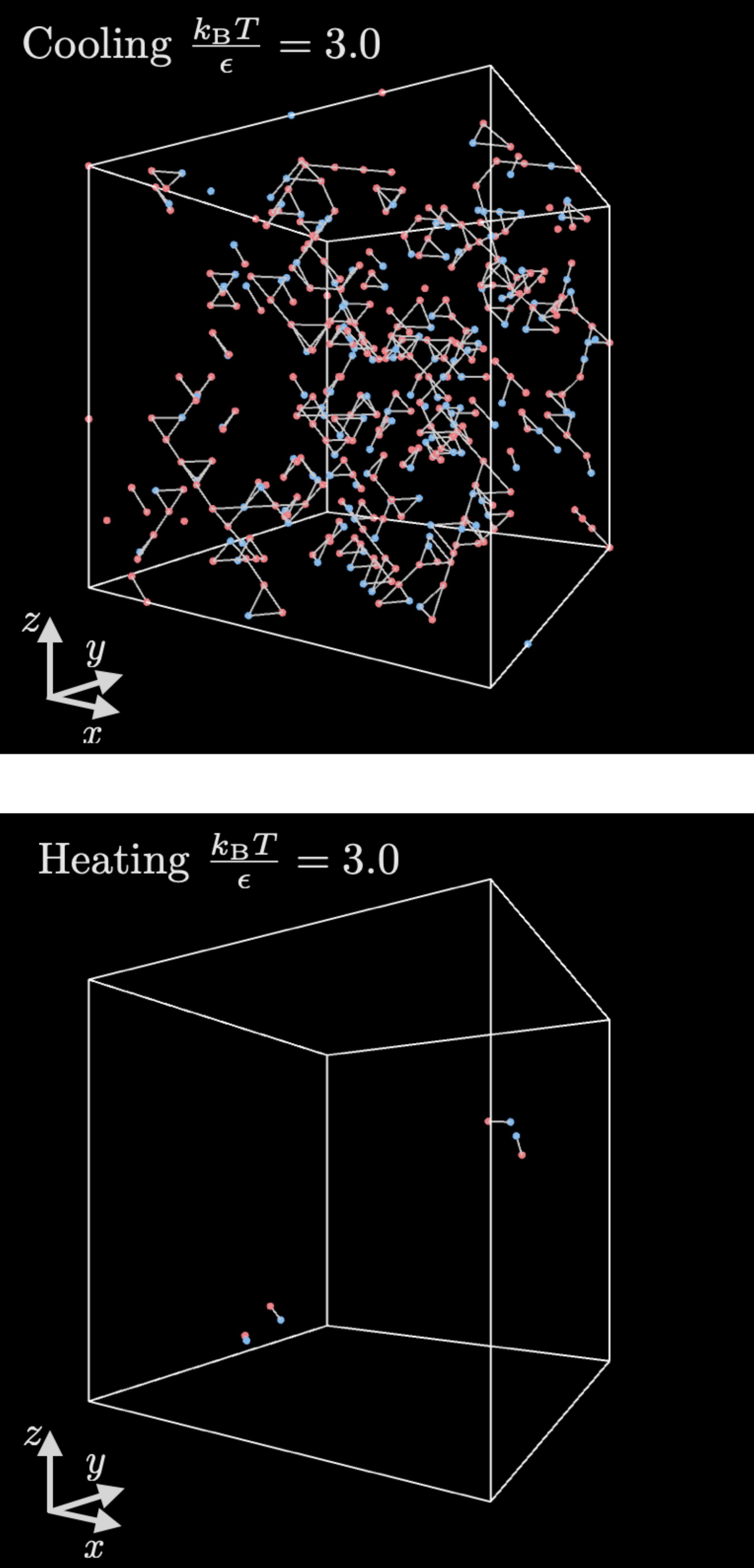}
\caption{
  Snapshots of the configuration of the system were observed in cooling [top] and heating [bottom] processes, corresponding to the supercooled liquid and crystalline states, respectively.
  The blue and red points represent Mo$^{4+}$ ions with Jahn-Teller energy
  $e_{\rm JT}^{\rm min}=-4\epsilon$ (type (\gtwo) in Table \ref{table1}.) 
  and $-16 \epsilon$ (type (\gfour) in Table \ref{table1}) respectively. 
  Otherwise, the Mo$^{4+}$ ions not shown have $e_{\rm JT}^{\rm min}=-12\epsilon$. 
  Here $L=9$ and $k_{\rm B}T/\epsilon=3.0$.
  Only the configuration inside $5 \times 5 \times 5$ cells are shown.
}
\label{mc_defect}
\end{figure}
\begin{figure}[t]
\includegraphics[clip,width=75mm]{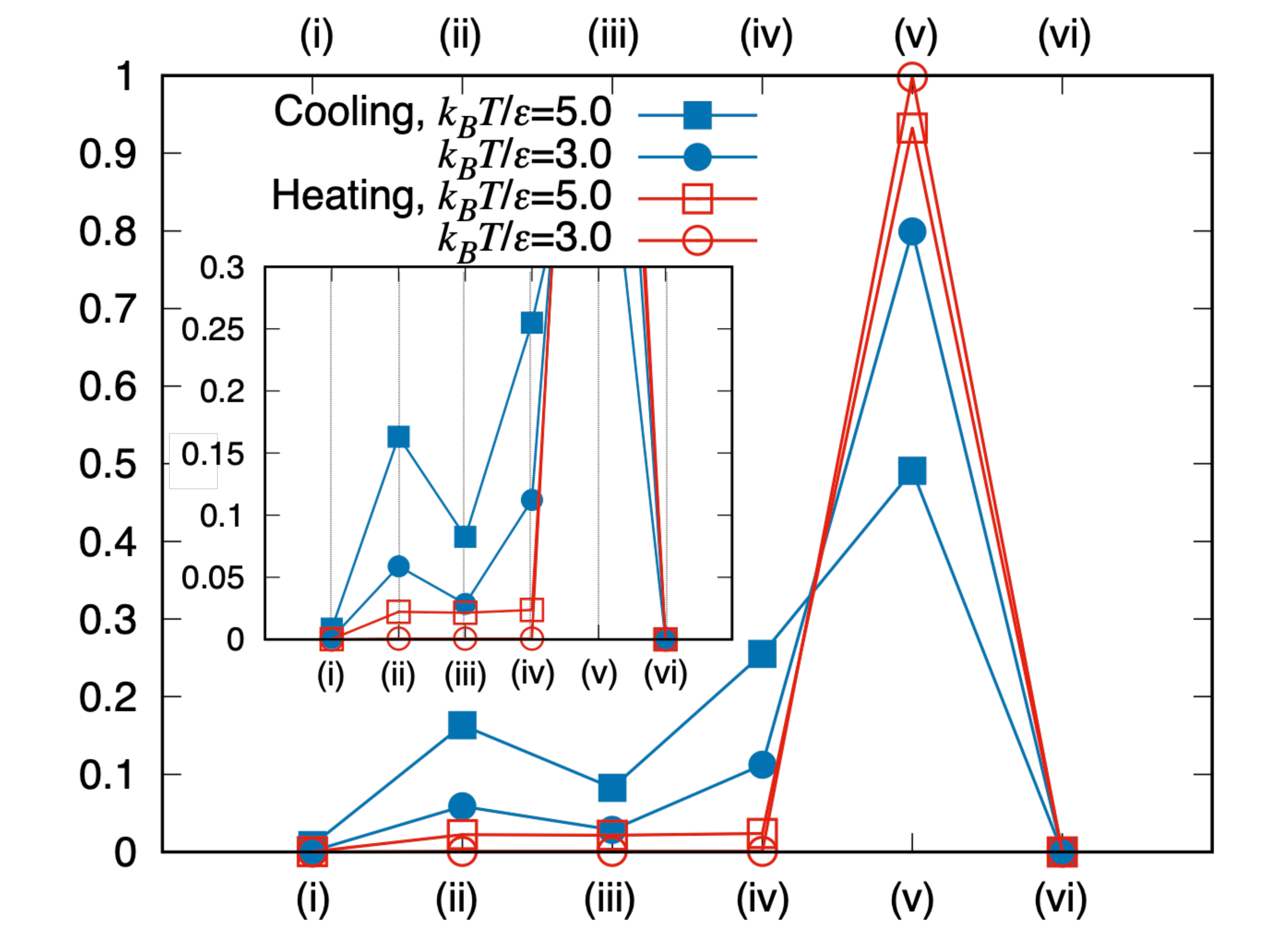}
\caption{
  The fraction of the displacement types (\gone)-(\gsix) in Table \ref{table1} 
  observed in the cooling and heating processes at $k_{\rm B}T/\epsilon=3.0$ and $5.0$ for $L=9$.
}
\label{pattern_fraction}
\end{figure}

Figure~\ref{mc_defect}(a) shows the snapshot of the supercooled liquid phase at $k_{\rm B}T/\epsilon=3.0$. 
The Mo$^{4+}$ ions with $e_{\rm JT}^{\rm min}=-12\epsilon$ having 2-$in$-2-$out$ tetrahedra on both sides 
are dominant which is not explicitly shown (see Table~\ref{table1}). 
Instead we visualize the Mo$^{4+}$ ions with non-2-$in$-2-$out$ structures: 
those with the lowest JT energy $e_{\rm JT}^{\rm min}=-16\epsilon$ in red and $-4\epsilon$ in blue. 
The former are surrounded by the latter and form disordered networks which fluctuate in time. 
Indeed, the two reds and one blue unit can be regarded as a trimer, which has an equivalent 
energy $-16\epsilon \times 2 -4\epsilon= -12\epsilon \times 3$ with the 2-$in$-2-$out$ ground state. 
However, generating such a trimer without the energy loss from the ground state is a very rare event. 
Figure~\ref{mc_defect}(b) shows the snapshot of the crystalline-like state in the heating process 
at $k_{\rm B}T/\epsilon=3.0$, where $P_{2-2}$ starts to deviate from 1. 
We only find a very few red-blue pairs of excitations while not the trimers. 
To quantitatively evaluate the nature of excitations, 
we plot in Fig.~\ref{pattern_fraction} the distribution of (\gone)-(\gsix) ions obtained 
by the Monte Carlo averages at different temperatures and heating/cooling processes. 
In the cooling process, the ratio of (\gtwo) to (\gfour) is overall 1:2, indicating that the trimer structure develops. 
At the same time, (\gthree) also contributes to the energy $-4\epsilon$, 
which indicates that the trimer structures are not perfectly kept 
which is the reason why the supercooled liquid has an energy density 
higher by $0.2\epsilon$ than the ground state $E= -6\epsilon$ (see Fig.~\ref{mc_sim}(d)). 

Contrastingly, in the heating process, the local structures other than (\gfive) start to be excited
only when we exceed the true first-order transition temperature. 
There, the density of (\gtwo), (\gthree), and (\gfour) do not differ much, 
and from the snapshots, it is confirmed not to originate from the trimers. 
\par
For these reasons, we consider that the 2-$in$-2-$out$ tetrahedra with a bending rule are required in the ground-state manifold, and exciting other patterns is a rare event that starts to happen when the temperature is increased up to the phase transition point. The supercooled liquid possesses a trimer-like structure but it accompanies several non-trimer structures. These excited structures fluctuate in the sea of 2-$in$-2-$out$, which contributes to a significantly large entropy of $\sim 0.05k_B$. 
However, eliminating these local structures requires high energy and the entropy of the ground state is possibly order-0, indicating that it is difficult to reach the ground state once we fall into the supercooled liquid phase. 

%
\subsection{Dynamic properties}
\begin{figure}[t]
\includegraphics[clip,width=85mm]{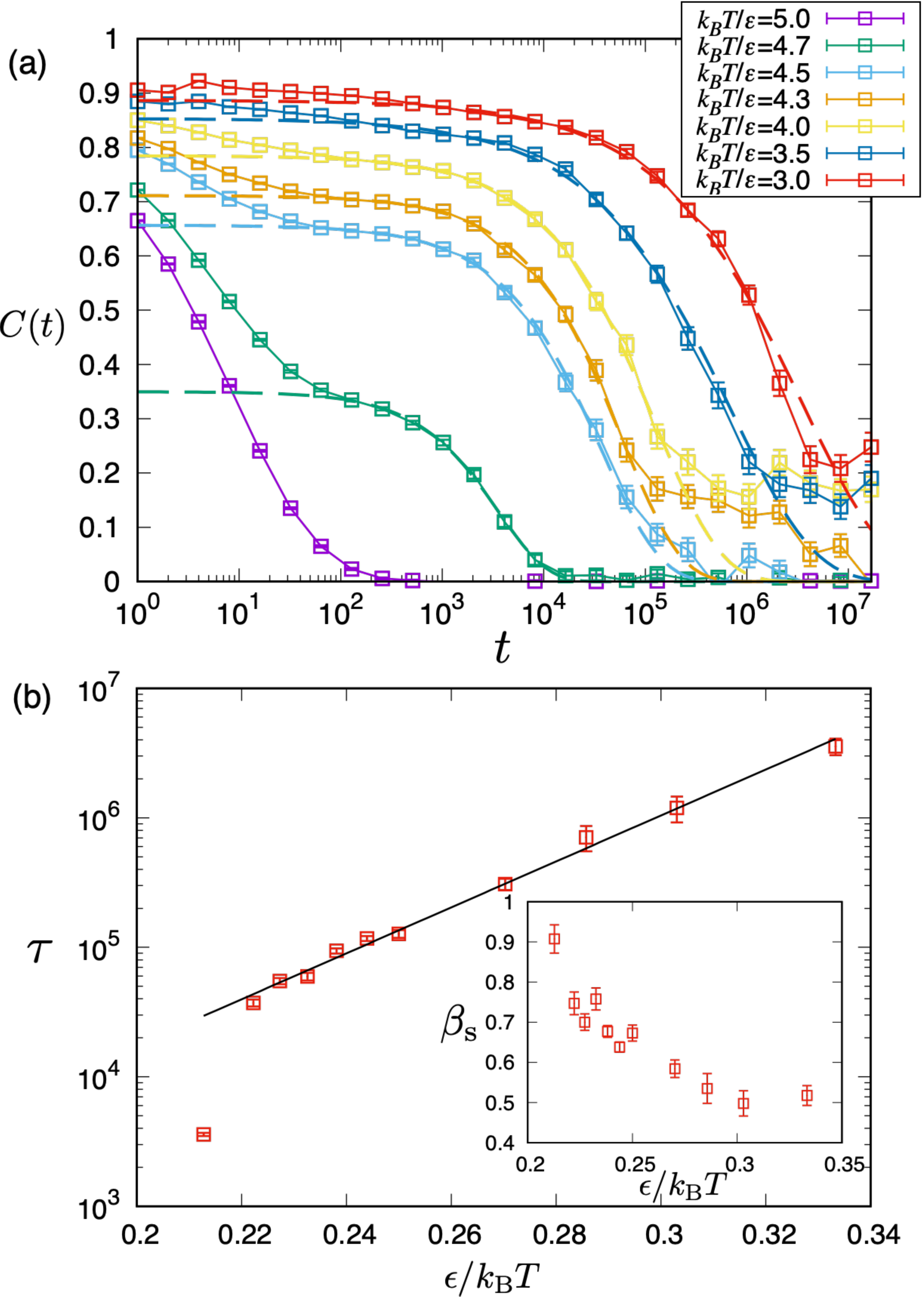}
\caption{ Relaxational dynamics in the supercooled liquid phase.
  (a) Auto-correlation function $C(t)$ for $L=9$ at various temperatures. 
  (b) Relaxation time $\tau$ and the stretching factor $\beta_{\rm s}$ (inset)
  obtained by fitting the data in panel (a) using \eq{eq-stretched-exponential}.
  The broken lines in (a) represent the fitted curves. 
  The inset shows the stretching factor $\beta_{\rm s}$ as a function of inverse temperature $\epsilon/k_{\rm B}T$. 
  }
\label{mc_auto}
\end{figure}

To understand further the nature of the supercooled liquid state, 
we analyze the relaxational dynamics in more detail. 
To this end, we measure the autocorrelation function of lattice displacement degrees 
of freedom $\sigma_i=\pm 1$ given as
\beq
C(t)=\frac{1}{N}\sum_{i=1}^{N} \langle \sigma_{i}(0)\sigma_{i}(t) \rangle
\eeq
where $t$ is the time measured in units of MCS. 
The initial configuration at $t=0$ is prepared in equilibrium state at $k_{\rm B}T/\epsilon=6.00$
similarly to the slow cooling protocol. 
Figure \ref{mc_auto} (a) shows $C(t)$ measured at different temperatures. 
At high enough temperatures, $k_{\rm B}T/\epsilon \sim 5.0$, $C(t)$ relaxes exponentially with time. 
However, on lowering the temperature the relaxation curve exhibits a plateau 
whose value shifts to higher positions for lower temperatures. 
Such two-step relaxation is a universal feature of supercooled glass-forming liquids \cite{angell2000relaxation}.
The first relaxation toward the plateau state is called $\beta$-relaxation, 
reflecting the short-time thermal fluctuation within the metastable states in which the system is temporarily trapped. 
Eventually, $C(t)$ leaves the plateau and starts to relax further, 
which is called $\alpha$-relaxation. 
\par
As shown in Fig.~\ref{mc_auto}(a), the $\alpha$ relaxation can be fitted by the stretched exponential form, 
\beq
C(t) \propto e^{-(t/\tau)^{\beta_{\rm s}}}.
\label{eq-stretched-exponential}
\eeq
and the temperature dependence of the relaxation time $\tau$ is well fitted by the Arrhenius law, 
\beq
\tau \propto e^{\frac{E_{\rm b}}{k_{\rm B}T}}
\label{eq-arrhenius}
\eeq
with a rather high energy barrier $E_{\rm b}/\epsilon=40.8 \pm 1.1$. 
The stretching factor $\beta_{\rm s}$ strongly depends on the temperature and decreases 
with lowering the temperature, as shown in the inset of Fig.~\ref{mc_auto}(a). 
Again, this stretched exponential decay is a universal feature observed in supercooled liquids
\cite{angell2000relaxation}.
If $\tau$ diverges at some temperature, it indicates a thermodynamic phase transition. 
However, in our case, $\tau$ follows a simple Arrhenius law in \eq{eq-arrhenius}, 
indicating that the system remains in the supercooled liquid state down to zero temperature 
with the indication of neither crystalline nor glass transitions. 
This kind of feature is often called a ``strong glass"  \cite{angell1991relaxation}.
\par
Closer inspection of  Fig.~\ref{mc_auto} (a) reveals the secondary plateau with a lower height 
that emerges at $k_{\rm B}T/\epsilon < 4.0$ at $t\gtrsim 1.0\times 10^5$. 
We leave the further investigation of this rich glassy dynamics for future investigations. 

\section{Discussion and Summary} 
\label{summary}
We analyzed theoretically the JT effect in a family of pyrochlore molybdates $A_2$Mo$_2$O$_7$ ($A=$Y, Dy, Tb) from a microscopic point of view. 
We examined in detail how the lattice distortions will influence both the elastic energy and the orbital energies in a crystal field through three processes. 
\par
First, we evaluated how the local on-site potential on the Mo$^{4+}$ ion generated from the surrounding six oxygen ions varies when it moves from the equilibrium position, finding that the $in$-$out$ displacement along the $Z$-axis is the softest among the three displacement modes. 
Next, by considering the elastic two-body interaction between the neighboring Mo$^{4+}$ ions, we showed that the 2-$in$-2-$out$ is the softest among all the vibrational modes of a single tetrahedron. 
Notice that this conclusion applies only to the $B$-site (Mo$^{4+}$ ion) of $A_2B_2$O$_7$ forming a pyrochlore lattice and neither to the $A$-site nor to the $B$-site of the $AB_2$O$_4$ with spinel structure. 
\par
In the third step, we examined the $e_g'$-orbital energy splitting of Mo$^{4+}$ ion induced by various $in$-$out$ lattice displacements, which changes the location of the surrounding six oxygen ions and modifies the crystal field. By examining all possible displacement patterns, we clarified that the displacement of Mo$^{4+}$ ions is energetically correlated over the third NN, namely between all ions belonging to adjacent tetrahedra. 
Among all these patterns, the 2-$in$-2-$out$ is the most favored, but unlike the standard spin ice Hamiltonian, not all the 2-$in$-2-$out$ join the lowest energy manifold, and there remain only limited numbers of 2-$in$-2-$out$ states which satisfy the {\it bending ice rule} in the ground state. 
\par
By introducing the Ising variables that represent the {\it in/out} lattice displacements of Mo$^{4+}$ ions, we derived the microscopic effective lattice Hamiltonian with two-body interactions between the nearest-neighbor, second NN, and third NN of the pyrochlore sites. This Hamiltonian faithfully reproduces the above-mentioned local JT energies. The existence of large second and third NN interactions implies a tougher constraint called the bending ice rule: among the 2-$in$-2-$out$ structures, those connecting the $in$-$in$ and $out$-$out$ bonds running in the same direction are excluded, whose phase space is significantly limited from the standard ice rule. 
\par
From the Monte Carlo simulations on the JT ice model, a more intriguing feature about the low-temperature state is clarified. In the cooling down, the system is trapped to the supercooled liquid state consisting of about 90\% of the 2-$in$-2-$out$ and 3-$in$-1-$out$/1-$in$-3-$out$ for the rest. This state is higher in energy than the ground state, and even though we take an extremely slow cooling rate, the first-order thermodynamic transition from the supercooled liquid to the ground state is avoided. Its dynamics is characterized by the two-step relaxation of the auto-correlation function of a typical supercooled liquid, and the timescale of the $\alpha$-relaxation follows a simple Arrhenius law, gradually slowing down toward zero temperature. We call this state a supercooled JT ice. 
\par
A very stable feature of the supercooled JT ice compared to the standard molecular supercooled liquid indicates two distinct features: there is a high free-energy barrier from the supercooled JT ice to enter a ground state. 
Both states have degeneracies while the former has much larger degeneracy, contributing to the entropy density of $\sim 0.05k_B$. 
These two features make a supercooled JT ice a good glass former, characterized by a metastable state that sustains down to zero temperature. 
\par
The lattice degrees of freedom are highly frustrated by themselves. 
However, its low energy structure has a notable difference from the standard spin ice or water ice. 
For the standard ice, the ground state has a residual entropy which amounts to about 30$\%$ of the total entropy, 
and the excitation takes place locally as a pair of monopoles, with its energy being extensive at the classical level. Therefore, the ice state gradually crosses over to the paramagnetic state at high temperatures. If an extra energy scale is added to the Hamiltonian, such as dipolar interactions\cite{melko2001long, fukazawa2002magnetic} or RKKY interactions\cite{nakatsuji2006metallic, ikeda2008ordering}, some of the states are selected as a ground state that has lower energy than the ice state, and the spin ice will undergo a first-order transition. Contrastingly, in the supercooled JT ice, even though there appears such ground state, the first-order transition from the metastable supercooled JT ice to the ground state is nearly prohibited due to a high energy barrier. It is known in the previous studies of spin ice systems that the dynamics of an ice-type model with further-neighbor interactions becomes slower than the one with only NN interactions \cite{rau2016spin, udagawa2016out}, in agreement with our observation. 
\red{However, in the spin ice, taking second-NN and third-NN interactions as comparable to or larger than the NN interaction is unphysical.
Therefore, our supercooled JT ice is a more distinct example of realizing such a situation in a natural manner.}
\par
Based on the present results, we now discuss the relevance of our supercooled JT ice with the glass phase. 
In several previous theories of pyrochlore magnets, the effect of lattice displacements is examined by assuming that they energetically favor a standard ice rule. For example, for the Ising-spin-lattice coupled Hamiltonian in Ref.[\onlinecite{smerald2019giant}], the 2-$in$-2-$out$ lattice displacements couple to the amplitude of the Ising spin-spin interactions, and yield a spin-lattice liquid state. 
\par
We have previously proposed the Heisenberg-spin and lattice coupled model and discovered the simultaneous spin-lattice glass transition without quenched randomness\cite{mitsumoto2020spin}. 
There, the lattice-lattice interactions of the first term in \eq{ene_all} are considered, and the spin-lattice coupling is microscopically derived in the form that the lattice displacement $\sigma_i$ changes the sign of the Heisenberg spin-spin interaction. At around the glass transition point, the 2-$in$-2-$out$ lattice displacements dominate the configuration of the lattice, i.e., $P\sim 0.9\text{-}1$. 
If we further take account of the second-NN and third-NN terms in \eq{ene_all}, the simple 2-$in$-2-$out$ lattice displacement is replaced by the supercooled JT ice, and we expect that the system will show a stronger tendency toward a glass transition. 
\par
The spin-glass material, Y$_2$Mo$_2$O$_7$, is a candidate of the supercooled JT ice, since the experiments show a variety of different types of lattice displacements of Mo$^{4+}$ ions which may possibly follow an ice rule. This material undergoes a spin-glass transition at 22K. 
The scale of the JT energy of this material is evaluated by substituting with Eqs. \eqref{epsilon} and \eqref{deltastar} the lattice constant $a \approx 2.03 ~(\AA)$, the amplitude of the JT distortion $\bar{\delta}^* \approx 0.121 ~(\AA)$, and the angle $\theta_0 = 61.76^\circ$, which are taken from Ref.[\onlinecite{thygesen2017orbital}]. 
The resultant JT energy per site is obtained as $12\epsilon/k_{\rm B} \approx 54.5$ (K). 
In our Monte Carlo simulation, the supercooled JT ice develops at around $k_B T\sim 4\epsilon$, 
namely at $10-30$K, where $P_{2-2}$ grows rapidly, 
which is consistent with the spin-glass transition 22K of the material. 
Below the transition temperature, the lattice displacement freezes but since this freezing is 
irregular, the crystal may safely keep its symmetry on an average. 
\par
There are some other pyrochlore materials whose lattice structure may show ice-like properties. 
MgTi$_2$O$_4$ is a pyrochlore material where Ti$^{3+}$ carries quantum spin-1/2 
and shows a 2-$in$-2-$out$ type of lattice displacement at 260 K \cite{isobe2002observation, schmidt2004spin}. 
However, the lower temperature phase is a valence bond crystal where the spins form a regular 2-$in$-2-$out$ dimerized state and the frustration is lost. 
In the tetragonal phase, where the material is located, the orbital degeneracy is absent, which means that in addition to the elastic property of the lattice the JT energetics of electrons is important for the realization of the JT ice. 
By a slight substitution of Mg ions to the nonmagnetic Ti ions, Mg$_{1+x}$Ti$_{2-x}$ enters a cubic phase, showing a suppression of susceptibility and an ice-like structural fluctuation\cite{torigoe2018nanoscale}. 
We may consider it as a recovery of JT activeness together with the 
introduction of site-randomness, and the cubic phase may be regarded as some sort of JT ice phase. 
Lu$_2$Mo$_2$O$_7$ is another candidate which shall be classified as one of the family members of 
the molybdates we focused on. 
This material has spin-1 carried by the two electrons on $4d$-orbitals, 
and shows a spin-glass behavior. 
However, for an oxynitride Lu$_2$Mo$_2$O$_5$N$_2$ with spin-1/2 the glass phase disappears and 
the resultant phase is regarded as a spin liquid\cite{clark2014spin}. 
Although the size of spins may play some role to explain the different behavior of the two materials, 
the disappearance of a glass phase may also be attributed to the lack of the JT effect, 
since we consider that the spin-glass transition is a cooperative transition of the 
JT lattice and the spin degrees of freedom. 
\par
The discovery of a good glassforming supercooled liquid phase of lattice degrees of freedom 
elucidated at the microscopic level is remarkable, 
because it can be a source of intrinsic disorder without quenched randomness in solids. 
This possibility shall be explored in future studies both in theories and in experiments. 
%
\begin{acknowledgments}
We thank Shunsuke Kito for discussions. 
This work is supported by JSPS KAKENHI (No. 19H01812). 
C.H. is supported by a Grant-in-Aid for Transformative Research Areas (Grant No. 21H05191)
and other JSPS KAKENHI (No. 21K03440, 18H01173) of Japan. 
\end{acknowledgments}
\appendix
\begin{figure}[t]
\includegraphics[clip,width=65mm]{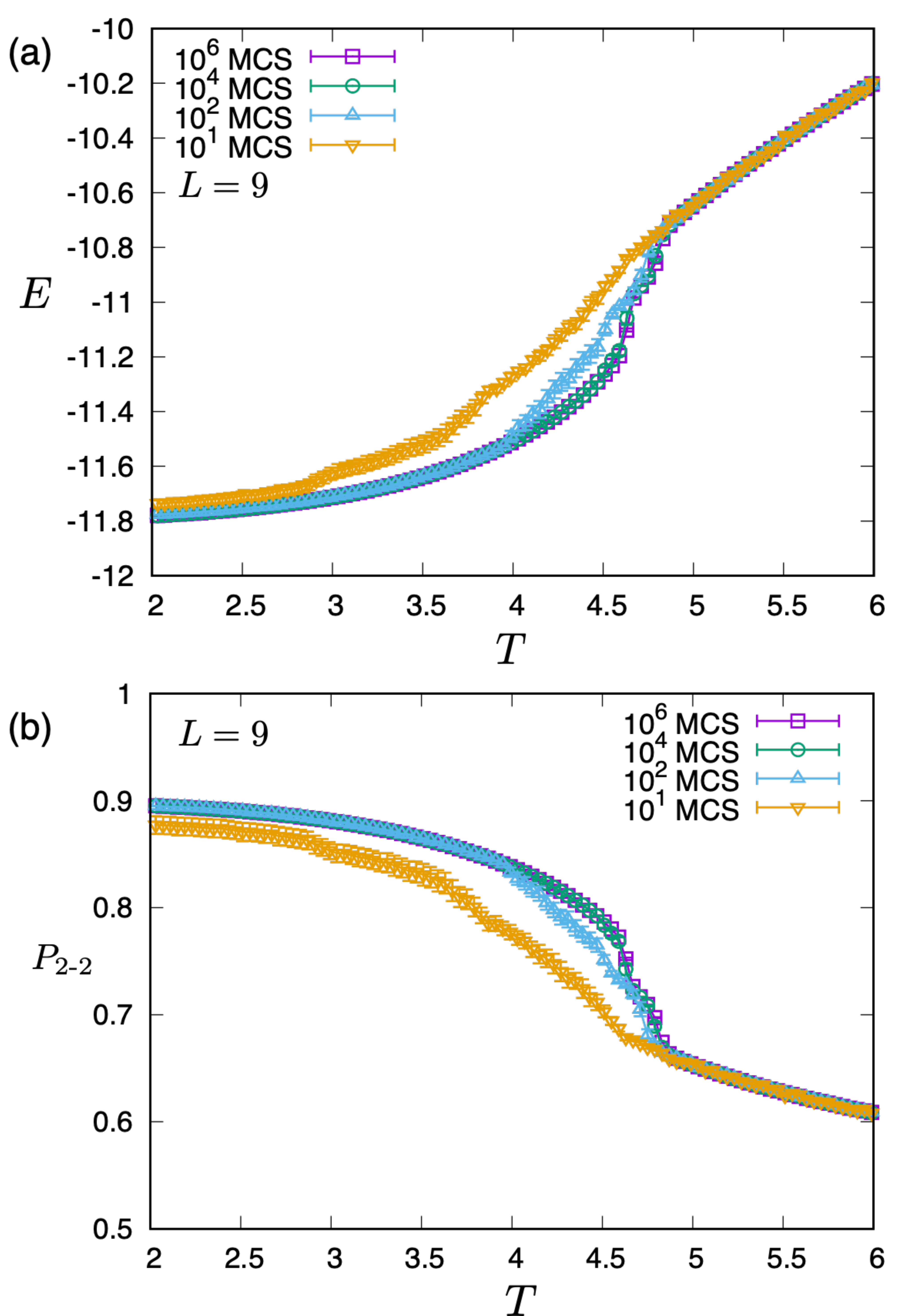}
\caption{Cooling rate dependences of (a) internal energy $E/\epsilon$ and 
(b) fraction of 2-in-2-out structure $P_{2-2}$. 
In all cases, the system is fully equilibrated initially at  $k_{\rm B}T/\epsilon = 6.00$
and then the temperature is lowered by a small step $\Delta (k_{\rm B}T/\epsilon) = 0.01$
where we take $\tau_{\rm cool}$ MCS both for equilibration and 
for taking thermal averages $\expval{\cdots}$ at each temperature. 
Here we display results for $\tau_{\rm cool}=1.0\times 10^1,10^{2},10^{3},10^{4},10^{6}$
with $L=9$. 
  }
\label{mc_cooling_rate_effect}
\end{figure}
\begin{figure}[t]
\includegraphics[clip,width=80mm]{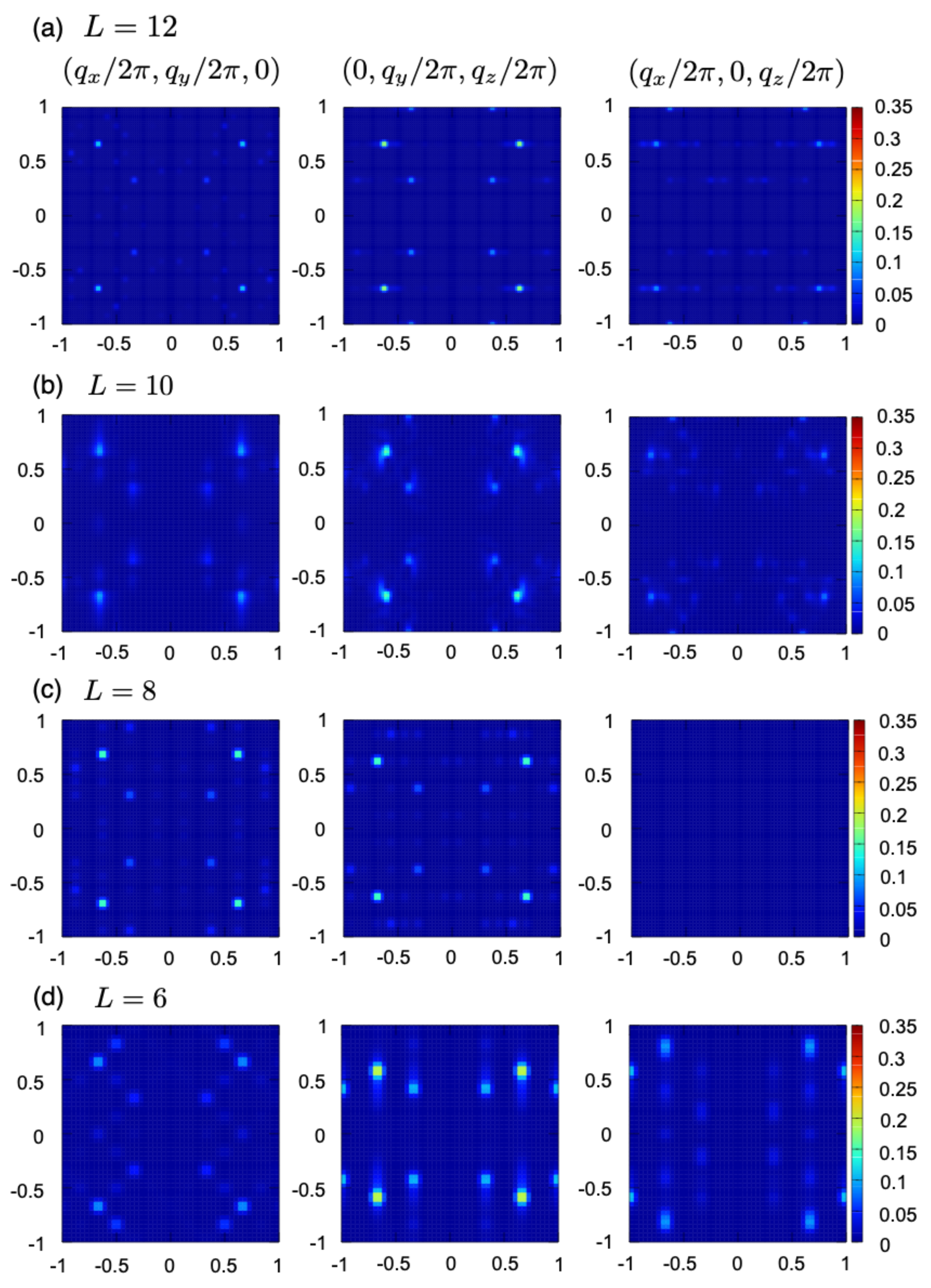}
\caption{
Density plot of the static structure factor $S_{\bm{q}}$ of the JT ice model
  obtained at $k_{\rm B}T/\epsilon = 3.0$ for several system sizes $L=6,8,10,12$ in the cooling process.
  We plot three slices in the $\bm q$-space as 
  $\bm q=(q_x/2\pi,q_y/2\pi,0)$, $(0,q_y/2\pi,q_z/2\pi)$ and $(q_x/2\pi,0,q_z/2\pi)$. 
  }
\label{structure_app}
\end{figure}
\section{Bending ice rule}
\label{sec-icerules}
We discuss the degree of degeneracy of the ground state of the JT ice model in Eq.(\ref{ene_all}), 
which follows a bending ice rule. 
First, we recall the estimation of the ground-state entropy introduced by Pauling 
for water ice or spin ice, which amounts to $S_{\rm water ice}= k_B\ln(3/2)^{N/2}$. 
This value coincides with the one obtained by the mean-field approximation which we 
expand in the following: 
a corner-sharing network of tetrahedra, including the pyrochlore lattice and checkerboard lattice, 
can be divided into two subgroups, i.e., dark and light-colored tetrahedra, 
as shown in Fig. \ref{fig_supercooling}(a). 
Only the tetrahedra with different colors are connected at each corner. 
If we assign 2-$in$-2-$out$ structures on all dark tetrahedra independently from each other, 
the number of states amounts to $6^{N/4}$. 
For a light tetrahedron connected to the four dark tetrahedra having $6^{4}$ choices of  2-$in$-2-$out$ structures, 
the probability of a light tetrahedron to also satisfy the 2-$in$-2-$out$ is $3/8$ on average. 
Hence, we can roughly estimate the number of the ground states as, 
\beq
W_{\rm water ice} = 6^{\frac{N}{4}}\qty(\frac{3}{8})^{\frac{N}{4}} = \qty(\frac{3}{2})^{\frac{N}{2}},
\eeq
and the entropy becomes
\beq
S_{\rm water ice}/N = \frac{k_{\rm B}}{N}\log\qty(\frac{3}{2})^{\frac{N}{2}} \approx 0.203 k_{\rm B}.
\eeq
It is known that the exact solution of a checkerboard ice is $S_{ice}= 0.216k_B$\cite{lieb1967} 
and the numerical evaluation of the pyrochlore ice is $S_{ice}= 0.205k_B$\cite{nagle1966}, 
which deviates only by 1$\%$.  
\par
Next, we employ this estimation to the JT ice model given in \eq{ene_all}. 
When we independently put 2-$in$-2-$out$ structures on all the dark tetrahedra, 
the probability that a light tetrahedron among four nearest-neighbor dark tetrahedra 
satisfies bending-rule mentioned in Sec. \ref{sec_simplify} is $2/27$. 
The number of the ground states then becomes 
\beq
W_{\rm JT ice} = 6^{\frac{N}{4}}\qty(\frac{2}{27})^{\frac{N}{4}} = \qty(\frac{2}{3})^{\frac{N}{2}}.
\eeq
However, this estimation is unphysical since the number $W_{\rm JT ice}$ decreases with increasing 
system size and approaches zero in the thermodynamic limit, which indicates a negative residual entropy. 
This suggests that the mean-field evaluation which assumes that there is an extensive 
degeneracy does not apply to the JT ice. 
In fact, we find that there are at least 12 states that satisfy the JT ice rule 
and have a periodic structure in a period of a unit cell. 
\par
Another way of considering the ice rule is roughly given as follows: 
let us first consider standard ice, and do not classify the color of red and blue bonds. 
For a given ice-configuration of a dark tetrahedron in Fig.~\ref{jti_gs_scl}(a) putting two bonds 
without connecting inside the tetrahedron, 
we find three different 2-$in$-2-$out$ configurations for each of the surrounding light tetrahedron. 
This means that we have $3^{N/2}$ configurations for uncolored bond connections 
which are exact for both the checkerboard and the pyrochlore lattice. 
However, we need to assign red and blue alternatively to these bonds, and the probability of 
having a proper connection for a single light tetrahedron is roughly 0.5 when the 
patterns of the surrounding four dark tetrahedra are determined. 
This will yield $W_{\rm water ice}=(3/2)^{N/2}$. 
In the same context, for the JT ice, we can assign a noncolored bond as $2^{N/2}$ following 
a bending rule. When assigning blue and red colors, we need to divide it by 2, 
and the resultant configuration number can be $W_{\rm JT ice}\sim (2/2)^{N/2}=1$. 
This may suggest that the entropy of the JT ice is zero, 
and $W_{\rm JT ice} \ge 12$ is of less than the order-$N$.

\section{High temperature expansion}
\label{sec-high-temperature-expansion}

To obtain the entropy in the high-temperature limit, we perform the high-temperature expansion for the JT ice model.
Supposing that the inverse temperature $\beta = \epsilon/k_{\rm B}T$ is small enough, i.e. $\beta \ll 1$, we can write the partition function up to the second-order with respect to $\beta$ as,
\begin{align}
\nonumber
Z &=\underset{\{\sigma_i \}}{\rm Tr} \exp(-\beta H) \\
&\approx \underset{\{\sigma_i \}}{\rm Tr} \qty[1 -\beta H + \frac{\beta^2 H^2}{2}],
\end{align}
where $\underset{\{\sigma_i \}}{\rm Tr}$ represents the sum of all microscopic states.
Using the relations, $\underset{\{\sigma_i \}}{\rm Tr} 1 = 2^N$, $\underset{\{\sigma_i \}}{\rm Tr} \sum_{i,j} \sigma_i \sigma_j = 0$ and $\underset{\{\sigma_i \}}{\rm Tr} (\sum_{i,j} \sigma_i \sigma_j)^2 = N_{\rm pair}2^N$ ($N_{\rm pair}$: total number of $(ij)$ pairs), we obtain 
\beq
Z \approx 2^N(1+15N\beta^2 \epsilon^2).
\eeq
The free energy per spin is calculated as
\beq
\frac{F}{N} \approx -\frac{1}{\beta} (\log 2 + 15\beta^2 \epsilon^2).
\eeq
Finally, we obtain the entropy in high temperature limit as
\beq
\frac{S}{N} = -\frac{1}{N} \dv{F}{T} = k_{\rm B}(\log 2 - 15\beta^2 \epsilon^2).
\eeq

\section{Cooling rate effects}
\label{sec-cooling-rate-effects}
In Figs.~\ref{mc_cooling_rate_effect}(a) and \ref{mc_cooling_rate_effect}(b) we show the internal energy and the fraction of 2-in-2-out structure obtained by different cooling rates. 
The cooling rate dependence appears at $k_BT\lesssim 4.8$. 
When we slow down the cooling rate, the energy decreases, and $P_{2-2}$ increases, 
while we find that the dependences of the cooling rate are well-converged 
when the equilibrating MCS is larger than $10^4$. 
\\
\\
\section{Size dependences of the structure factors}
\label{sec-size dependences}
In Fig.~\ref{structure_app} we show the static structure factor several system sizes $L=6,8,10,12$ in the supercooled liquid state. 
The peaks near $(\pm 4\pi/3,\pm 4\pi/3,0)$ or $(0,\pm 4\pi/3,\pm 4\pi/3)$ 
but are slightly off these points are observed for different $L$, 
which corresponds to the nearly three-fold periodic short-range ordering. 
However, the precise peak positions differ for different sizes, indicating that 
the size and the periodic boundary conditions of a finite size lattice pins 
the most favorable structures. 
This result indicates that there is an underlying competition of different orders 
in the supercooled JT ice phase. 

\bibliography{jti_ref}

\end{document}